\documentclass[11pt,a4paper]{article}
\pdfoutput=1
\usepackage{microtype}
\DisableLigatures[f]{encoding = *, family = * }
\usepackage{upref,amsmath,amssymb,amsthm,mathrsfs,epsf,amsfonts}
\usepackage{graphicx}
\usepackage{hyperref}
\usepackage{paralist}
\usepackage{multirow}
\usepackage{float}
\usepackage{url}
\usepackage{rotating}
\usepackage{epstopdf}
\usepackage{tablefootnote}
\usepackage{xfrac}
\usepackage{paralist}
\usepackage[round]{natbib}
\usepackage{caption}
\usepackage{subcaption}
\usepackage[table]{xcolor}
\usepackage[version=0.96]{pgf}
\usepackage{tikz}
\usetikzlibrary{arrows,shapes,snakes,automata,backgrounds,petri}
\usepackage{pgfplots}
\usepackage{authblk}
\usepackage[section]{placeins}

\usepackage{setspace}
\newcommand*{\equal}{=}
\newcommand{\trsp}{\ensuremath{\top}}
\newcommand{\mv}[1]{{\boldsymbol{#1}}}
\renewcommand{\det}[1]{\left\lvert#1\right\rvert}
\DeclareMathOperator{\trace}{tr}

\DeclareMathOperator{\vectorize}{vec}
\newcommand*{\md}{\,\mathrm{d}}
\newcommand*{\bmat}[1]{%
\begin{bmatrix}#1\end{bmatrix}}

\newcommand*{\R}{\mathbb{R}}

\newcommand*{\proper}{\mathsf}

\newcommand*{\pE}{\proper{E}}

\graphicspath{{./Figure/}}

\begin{document}
\title{\bf Modelling Spatial Compositional Data:\\ \large{Reconstructions of past land cover and uncertainties}}
\author[1,2]{Behnaz Pirzamanbein \thanks{This research is part of two Swedish strategic research areas: Biodiversity and Ecosystems in a Changing Climate (BECC) and ModElling the Regional and Global Earth system (MERGE). The paper is also a contribution to PAGES LandCover6k.
Poska received financial support from FORMAS LUsTT project and ETF project 9031.
Gaillard received financial support from the Faculty of Health  and Life Sciences, Linnaeus University, Kalmar, Sweden.
}
}
\affil[1]{\small Centre for Mathematical Sciences, Lund University, Sweden}
\affil[2]{Centre for Environmental and Climate Research, Lund University, Sweden}
\author[1]{Johan Lindstr\"{o}m}
\author[3,4]{Anneli Poska}
\affil[3]{Department of Physical Geography and Ecosystems Analysis, Lund University, Sweden}
\affil[4]{Institute of Geology, Tallinn University of Technology, Estonia}
\author[5]{Marie-Jos\'{e} Gaillard}
\affil[5]{Department of Biology and Environmental Sciences, Linnaeus University, Sweden}
\renewcommand\Authands{ and }
\date{\vspace{-5ex}}
\maketitle

\begin{abstract}
In this paper, we construct a hierarchical model for spatial compositional data, which is used to reconstruct past land-cover compositions (in terms of coniferous forest, broadleaved forest, and unforested/open land) for five time periods during the past $6\,000$ years over Europe. The model consists of a Gaussian Markov Random Field (GMRF) with Dirichlet observations. A block updated Markov chain Monte Carlo (MCMC), including an adaptive Metropolis adjusted Langevin step, is used to estimate model parameters. The sparse precision matrix in the GMRF provides computational advantages leading to a fast MCMC algorithm. Reconstructions are obtained by combining pollen-based estimates of vegetation cover at a limited number of locations with scenarios of past deforestation and output from a dynamic vegetation model. To evaluate uncertainties in the predictions a novel way of constructing joint confidence regions for the entire composition at each prediction location is proposed. The hierarchical model's ability to reconstruct past land cover is evaluated through cross validation for all time periods, and by comparing reconstructions for the recent past to a present day European forest map. The evaluation results are promising and the model is able to capture known structures in past land-cover compositions.
\end{abstract}
 
\noindent%
\vfill

\newpage

\section{Introduction}
\label{sec:intro}
Modelling the spatial distribution in species composition and the relative
abundances of different species is a common problem in environmental studies.
\citep{aitchison1986statistical, paciorek2009mapping, billheimer2001statistical,
  tjelmeland2003composition, Pirzamanbein2014}. In this paper we develop a
statistical model for spatial compositional data and a way of assessing the
uncertainties in the resulting compositional reconstructions at unobserved
locations. 
The model is used to reconstruct past land-cover composition over Europe from local pollen-based estimates of vegetation cover.

\subsection{Spatial Interpolation of Compositional Data} 
A common approach to modelling compositional data is Gaussian modelling of log-ratio
transformed data \citep{aitchison1986statistical}, where the spatial structure
can be captured using Gaussian fields \citep{billheimer2001statistical,
  tjelmeland2003composition, Pirzamanbein2014}. However, modelling transformed
compositions as Gaussian might understate the uncertainty in the data,
especially in cases of over-dispersion \citep{paciorek2009mapping}.

To capture the variability in our observations, we propose a Bayesian
hierarchical model (described in Sec.~\ref{sec:model}) where the
compositional data are seen as Dirichlet observations of an underlying latent
field of probabilities. 
The field of compositional probabilities is in turn modelled using a transformed Gaussian Markov Random Field (GMRF) \citep{rue2004gaussian, lindgren2011explicit}. The sparsity in the precision matrix of the GMRF allows us to compute the Hessian for the entire latent field, allowing for fast estimation (see Sec.~\ref{sec:MCMC}) using a Metropolis Adjusted Langevin algorithm (MALA) \citep{MALA,RoberS2003_4}. 

To describe the uncertainties in the compositional reconstructions we propose a novel way of computing joint confidence and prediction regions for compositional data (Sec.~\ref{subsec:Uncertainty}). The method accounts for the interdependence among the components of the compositional data and allows us to illustrate the joint uncertainty in the composition at each prediction location.


\subsection{Climate Studies and Past Land Cover}
For climate modelling studies the land-cover composition is commonly divided into three land cover types: coniferous forest, broadleaved forest, and unforested/open land. The spatial distribution of these land cover types play an important role in the climate system \citep{claussen2001}. Accurate, spatially continuous, descriptions of past land cover types are necessary to assess past land cover-climate interactions \citep{borovkin2006} and the impact of anthropogenic land-cover changes on climate \citep{strandberg2014, gaillard2010holocene, team2008assessment,Pirzamanbein2014}.

Historic maps and surveys of past land cover have limited temporal coverage
(rarely more than the past 300 to 500 years) and is often spatially fragmented
due to a lack of transnational databases. Land-cover in climate
models is currently implemented using a combination of dynamic vegetation models
\citep[e.g. LPJ-GUESS][]{smith2001representation} and scenarios of
anthropogenic land-cover changes \citep[e.g.][]{kaplan2009prehistoric,
  klein2011hyde, pongratz2008}. Here, the dynamic vegetation models provide a climate-induced, potential vegetation, which is modified by the anthropogenic scenarios to account for human activities (mainly deforestation).

Land-cover reconstruction from fossil pollen records is an alternative, to
the dynamic vegetation model simulations and anthropogenic scenarios, that may provide more realistic descriptions of past land cover for climate modelling studies \citep{gaillard2010holocene, trondman2015}.
Given pollen records extracted from lakes and bogs, pollen-based estimates
of vegetation cover are obtained using a model \citep[here the REVEALS
model of][]{sugita2007Love, sugita2007Reveals}. The model provides estimates of 
pollen-based land-cover composition (hereafter called PbLCC) for a limited area (ca. 100 km x 100 km) around each lake or bog. For use in climate modelling these PbLCC estimates need to be interpolated into continuous maps of past land-cover composition at sub-continental to global scales \citep{Pirzamanbein2014, paciorek2009mapping}.

The PbLCC data used in this paper are available for five time periods during the
past $6\,000$ years, and the proposed Bayesian model and estimation procedure is used to interpolate the PbLCC data for each time period. The results are validated using present-time forest maps and cross-validation (Sec.~\ref{subsec:validation}). The model shows good predictive power, capturing
known structures and historical changes in land-cover composition. 

The paper ends with some brief conclusion in Sec.~\ref{sec:conclusion}.

\section{Model}
\label{sec:model}
To model the spatial structure in the compositional data we propose a hierarchical model, where the observed compositions at each location are modelled as draws from a Dirichlet distribution. The Dirichlet is parametrized using a scale (or concentration) parameter and a vector of probabilities. The spatial dependence in these compositional probabilities is modelled using a transformed GMRF. Details regarding the observational model are given in Sec.~\ref{sec:Dir}, and Sec.~\ref{sec:eta} describes the latent field.

\subsection{Dirichlet Distribution and Link Function}
\label{sec:Dir}
Compositional data are discussed in detail by \citet{aitchison1986statistical}, here a brief overview is given. Let $\mv{y}_s=(y_{s,1}, y_{s,2}, \cdots, y_{s,D})$ be the $D$-compositional data at location $u_s\in \mathbb{R}^2, \; s=1, \cdots, N_o$, the restrictions for compositional data imply that: $y_{s,k} \in (0,1)$ and $\sum_{k=1}^D y_{s,k} = 1$. Conditional on the transformed underlying field, $\mv{z}=f(\mv{\eta})$, we assume that the data, $\mv{Y}=\{\mv{y_s}\}_{s=1}^{N_o}$, are independent draws from a multivariate Dirichlet distribution, 
\begin{equation}
\mathbb{P}(\mv{Y}|\alpha,\mv{z}) = \prod_{s=1}^{N_o} \Big( \dfrac{\Gamma(\alpha)}{\prod_{k=1}^D \Gamma(\alpha z_{s,k})}\prod_{k=1}^D y_{s,k}^{\alpha z_{s,k}-1}\Big), \quad \alpha >0,
\label{eq:dirichlet}
\end{equation}
and
\begin{align*}
z_{s,k} &\in (0, 1), & \sum_{k=1}^D z_{s,k} &= 1
\end{align*}
where $\alpha$ is a Dirichlet scale parameter. 

The link function, $f$, between $\mv{z}$ and $\mv{\eta}$ can be any function
from $\mathbb{R}^{d \times N_o}$ to $(0, 1)^{D \times N_o}$ such that:
\begin{equation}
  \begin{split}
    f(\eta_1, \cdots, \eta_d) &= (Z_1,\cdots,Z_d,Z_D),\\
    \sum_{k=1}^D Z_{k} &= 1, \quad \text{and} \quad d=D-1.
\end{split}
\label{eq:link}
\end{equation}
Here $Z_k$ is a $N_o\times1$ column vector containing the $k^\text{th}$
component of the D-compositional data and $\eta_k$ is a column vector with the
$k^\text{th}$ latent field, i.e. the probabilities and latent 
fields for location $s$ are given by $\{z_{s,k}\}_{k=1}^D$ and $\{\eta_{s,k}\}_{k=1}^d$,  respectively.

In this paper the link function is constructed by applying the additive
log-ratio \citep{aitchison1986statistical} transform 
\begin{equation}
  \eta_{s,k} = \log z_{s,k} - \log z_{s,D}, \quad k=1,\ldots,D-1
  \label{eq:alr}
\end{equation} 
for each location $s$. The possible choices of transforms at each location also
include the isometric log-ratio transformation \citep{EgozcPMB2003_35}. However,
it excludes the central log-ratio transformation
\citep{aitchison1986statistical}, which gives a latent $\mv{\eta}$-field with
unidentifiable mean.

\subsection{Latent Field}
\label{sec:eta}
Given a total of $N \geq N_o$ locations at which we want to provide composition
predictions the latent field, $\mv{\eta}_{all}$, is multivariate with $d=D-1$
elements at each location ($N \geq N_o$ since we are providing predictions at
the observed and additional locations). To simplify notation the latent
field is represented as a $Nd \times 1$ vector $\mv{\eta}_{all} =
({\eta}_{all,1}^\trsp,\cdots,{\eta}_{all,d}^\trsp)^\trsp$, 
where each $\eta_{all,k}$ is spatial field with $N$ locations.

The latent field and its connection to the observed locations is given as:
\begin{equation}
\begin{split}
\mv{\eta} &= \mv{A}\mv{\eta}_{all}\\
\mv{\eta}_{all} &= \mv{B}\mv{\beta}+ \mv{X}.
\end{split}
\label{eq:eta}
\end{equation}
where $ \mv{A}= \mathbb{I}_{d\times d}\otimes A$ extracts the observed elements from $\mv{\eta}_{all}$, with $A$ being a $N_o \times N$ sparse observation matrix; $\mv{B} = \mathbb{I}_{d\times d}\otimes B$ with $B$ being a $N\times p$ matrix of covariates; $\mv{\beta}$ is a $dp\times 1$ matrix of regression coefficients; and $\mv{X} = (X_1^\trsp,\cdots,X_d^\trsp)^\trsp$ is a spatially correlated multivariate field. With this structure, the spatial dependence $\mv{X}$, can be modelled as a GMRF with a separable covariance structure, i.e. $\mv{\rho}\otimes \mv{Q}^{-1}$, which captures the dependency among and within the fields;
\begin{equation}
\mv{X} \sim \mathcal{N}(\mv{0}, \mv{\rho}\otimes \mv{Q}^{-1}(\kappa)).
\label{eq:X}
\end{equation}
Here $\mv{\rho}$ is a $d\times d$ matrix of covariances among the $d$
multivariate fields ($X_k, \, k=1, \cdots, d$), and $\mv{Q}(\kappa)$ is a
$N\times N$ precision matrix of a GMRF with spatial scale parameter $\kappa$.
$\mv{Q}$ is chosen as a precision matrix which approximates a stationary
Mat{\'e}rn field \citep{Mater1960_Spatial} with smoothness $\nu=1$;
\begin{align}
  \mv{Q}(\kappa) 
  =\kappa^4 \mv{C} + 2 \kappa^2 \mv{G} + \mv{G}\mv{C}^{-1}\mv{G}.
\label{eq:Q}
\end{align}
Here $\mv{C}$ is a diagonal matrix and $\mv{G}$ is a finite difference
approximation of the negative Laplacian \citep[cf.\ Appendix A in ][]{lindgren2011explicit}.
This precision is also a solution to a stationary stochastic partial deferential
equation (SPDE) field with $\alpha = 2$ \citep[see ][ for
details]{lindgren2011explicit}. While the smoothness, $\nu$, of the latent field
is known to affect spatial prediction \citep{Stein1999} it is also very hard to
estimate \citep{Haran2011_Brooks} and a popular default is to use the
exponential covariance ($\nu=0.5$). For GMRF models in $\R^2$, $\nu$
has to be integer resulting in our choice of $\nu=1$; a value also suggested by
\citet{Whitt1954_41}. It should further be noted that for $\nu=1$ (or $\alpha=2$)
the special-case of $\kappa=0$ (infinite range) gives \citet{Wahba1981_2}
splines, providing a link between spline smoothing and Gaussian
spatial-processes \citep{KimelW1970_41, Nychk2000_Schimek}.

\subsection{Hierarchical Model and Priors}
\label{sec:hier}
\begin{figure}
\centering
\begin{tikzpicture}[node distance=1cm,>=stealth',bend angle=45,auto]
  \tikzstyle{place}=[circle,thick,draw=blue!75,fill=blue!20,minimum size=5mm]
  \tikzstyle{transition}=[rectangle,thick,draw=black!75,fill=black!20,minimum size=5mm]
  \tikzstyle{trans}=[ellipse,thick,draw=gray!75,fill=gray!20,minimum size=6mm]
  \tikzstyle{transitionw}=[rectangle,thick,draw=gray!75,fill=gray!20,minimum size=5mm]
  \tikzstyle{place_eta}=[ellipse,thick,draw=blue!75,fill=blue!20,minimum size=3mm]
  \begin{scope}
    
    \node [transition] (a)  {$a_\kappa$};
    \node [transition] (b)  [right of=a] {$b_\kappa$};
    \node [transition] (df) [right of=b] {$df$};
    \node [transition] (I)  [right of=df] {$\mathbb{I}$};

    \node [place] (kappa) [below of=a, xshift=10mm]  {$\kappa$}
    edge [pre] (a)
    edge [pre] (b);
    \node [place] (rho)   [below of=df, xshift=5mm]  {$\mv{\rho}$}
    edge [pre] (df)
    edge [pre] (I);

	\node [transition] (q)  [left of=kappa] {$q_\beta$};
	
	\node [transition] (aalpha)  [right of=rho,xshift=5mm] {$a_\alpha$};
    \node [transition] (balpha)  [right of=aalpha] {$b_\alpha$};
	
	\node [place_eta] (eta)   [below of=kappa,xshift=5mm,yshift=-2mm] {$\mv{\eta}_{all}=\mv{B}\mv{\beta}+ \mv{X}$}
	edge [pre] (kappa)
    edge [pre] (rho)
    edge [pre] (q);
    
    \node [place] (alpha)   [below of=aalpha,xshift=3mm,yshift=-2mm] {$\alpha$}
	edge [pre] (aalpha)
    edge [pre] (balpha);
    
	\node [trans] (z)  [below of=eta,yshift=-5mm]{$\mv{z} \equal f(\mv{A}\mv{\eta}_{all})$}
          	edge [pre] (eta);
     
	\node [transition] (y)  [below of=z,xshift=18mm,yshift=-2mm] {$\mv{Y}$} 
    edge [pre] (alpha)
    edge [pre] (z);
  \end{scope}
\end{tikzpicture}
\caption{Directed acyclic graph describing the conditional dependencies on the hierarchical model.}
\label{fig:hierarchical}
\end{figure}
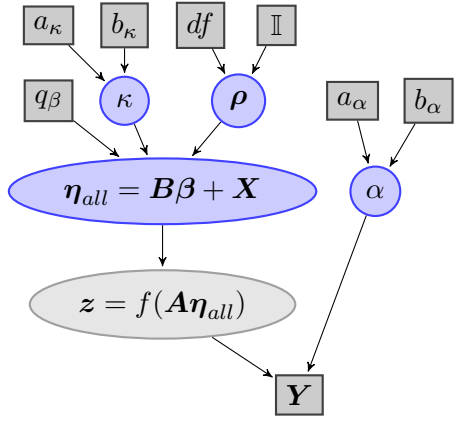

The full hierarchical model (Fig. \ref{fig:hierarchical}) based on Dirichlet observations \eqref{eq:dirichlet} of a transformed latent GMRF \eqref{eq:X} becomes
\begin{equation}
	\begin{aligned}
		\mv{y_s}|\alpha, \mv{\eta} &\sim Dir(\alpha f_s(\mv{\eta})), & s&=1,\ldots,N_o
    \\
		\mv{\eta}_{all} &= \mv{B}\mv{\beta}+ \mv{X}, & \mv{\eta}&= \mv{A}\mv{\eta}_{all},\\
		\mv{X}|\kappa, \mv{\rho} &\sim \mathcal{N}(\mv{0}, \mv{\rho}\otimes \mv{Q}^{-1}(\kappa)),\\
		\mv{\beta} &\sim \mathcal{N}(0,\mathbb{I}q^{-1}_\beta), & \alpha &\sim \Gamma(a_\alpha, b_\alpha)\\
		\kappa &\sim \Gamma(a_\kappa, b_\kappa), & \mv{\rho}|\kappa &\sim IW(a_\rho \mathbb{I}, b_\rho)
  \end{aligned}
  \label{eq:full}
\end{equation}
where $\mathbb{I}$ are appropriate identity matrices. To make $\mv{X}$ and $\mv{\beta}$ jointly normal, we use a vague Gaussian prior for $\mv{\beta}$ with precision $q_\beta =10^{-3}$. The Dirichlet scale parameter, $\alpha$, and spatial scale parameter, $\kappa$, are given gamma priors, and for $\mv{\rho}$ we choose a conjugate prior for covariance matrices, the inverse Wishart ($IW$). The
conjugacy of the inverse Wishart provides computational advantages when updating
the parameters of our multivariate latent field.

A suitable prior on $\kappa$ can be obtained by noting its link
with the range of the field: $\text{range} \approx \sqrt{8\nu} / \kappa$
\citep{lindgren2011explicit}. This link is used by R-INLA to create reasonable
defualt priors where the range is related to the size of the domain
\citep{LindgR2015_63}. An alternative option is presented by
\citet{FuglsSLR2016_Interpretable} and \citet{SimpsRMRS2015_Penalising}, introducing a
prior that shrinks towards $\kappa=0$, the intrinsic field (i.e.\ a spline 
smoother). This prior is motivated by the intrinsic field representing a simpler
model, to be prefered in the absence of convincing data. The prior in
\citet{FuglsSLR2016_Interpretable} corresponds to $a_\kappa=1$ with $b_\kappa$
chosen to give a suitably small prior-probability to short ranges. A $1\%$
probability of $\text{range}<1$ results in $b_\kappa = -\log(0.01) \cdot 1 
/ \sqrt{8}$, the range of $1$ is based on the unit distance between
our gridcell centroids. 
For the inverse Wishart prior on $\mv{\rho}$, we chose uninformative prior with $a_\rho = 1$ and $b_\rho = 10$. The inverse Wishart is proper if the degree of freedom is $b_\rho > d-1$, has finite mean if $b_\rho > d+1$ and has finite variance if $b_\rho > d+3$. In practice $b_\rho$ is often chosen somewhat larger than these lower bounds \cite[see e.g. ][]{schmidt2010dof}. 
Given our lack of intuition for $\alpha$ we pick uninformative prior resulting in the following values of all hyper parameters:
\begin{align*}
 a_\alpha &= 1.5, & a_\kappa &= 1, & a_\rho &= 1,\\
 b_\alpha &= 0.1, & b_\kappa &= \frac{\log(100)}{\sqrt{8}}, &
   b_\rho &= 10.\\
\end{align*}

Having detailed the model, parameter estimation and reconstruction of
$\mv{\eta_{all}}$, using MCMC, are described in the following section.

\section{Estimation Using MCMC}
\label{sec:MCMC}
A block-updated MCMC algorithm is used to estimate the latent field $\mv{\eta}_{all}$ and the unknown parameters $\alpha, \kappa, \mv{\rho}$. For GMRF, joint updating of parameters in as large blocks as possible has been shown to improve mixing and convergence \citep{BlockingRueHeld}. Therefore, the algorithm in this paper updates the unknowns by alternating between two blocks: the first block updates the latent fields and the Dirichlet scale parameter using MALA \citep{MALA,RoberS2003_4}, the second block updates the parameters of the GMRF, $\kappa$ and  $\mv{\rho}$, using a combination of random walk proposals and the conjugate posterior for $\mv{\rho}$.

\subsection{Updating $\mv{\eta}_{all}$ and $\alpha$}
To update $\mv{\eta}_{all} = \mv{B}\mv{\beta}+ \mv{X}$ and $\alpha$ we use a Metropolis-Hastings step to draw samples from the conditional distribution
\begin{equation}
\begin{split}
\mathbb{P}(\mv{X},\mv{\beta}, \alpha|\kappa,\mv{\rho},\mv{Y}) \propto& \big(\prod_{s=1}^{N_o} \mathbb{P}(\mv{y}_s|f_s\big(\mv{A}\mv{\eta}_{all}\big),\alpha\big) \cdot \mathbb{P}(\mv{X}|\kappa,\mv{\rho}) \cdot \mathbb{P}(\mv{\beta}) \cdot \mathbb{P}(\alpha)\\
\propto& \prod_{s=1}^{N_o} (\dfrac{\Gamma(\alpha)}{\prod_{k=1}^D \Gamma(\alpha z_{s,k})}\prod_{k=1}^D y_{s,k}^{\alpha z_{s,k}-1})\\
&\cdot \exp \big(-\frac{1}{2} \mv{X}^{\trsp} \big(\mv{\rho}^{-1}\otimes \mv{Q}(\kappa)\big) \mv{X} \big)\\
&\cdot \exp \big(-\frac{q_{\mv{\beta}}}{2} \mv{\beta}^{\trsp} \mv{\beta} \big) \cdot \alpha^{a_\alpha-1}e^{-\alpha\cdot b_\alpha}.
\label{eq:post_eta_alpha}
\end{split}
\end{equation}
The Metropolis-Hastings step uses a MALA proposal:
\begin{equation}
\mv{X}^*,\mv{\beta}^*, \alpha^* | \mv{X},\mv{\beta}, \alpha \quad \sim \quad \mathcal{N} \left(
(\mv{X},\mv{\beta}, \alpha)^\trsp
+ {\epsilon^2 \over 2} \mv{\mathcal{I}}^{-1} \nabla l ,  \epsilon^2 \ {\mv{\mathcal{I}}}^{-1}\right),
\label{eq:MALA}
\end{equation}
where $\epsilon$ is the step size of MALA, $\nabla l$ is a vector of derivatives
of $\log \mathbb{P}(\mv{X},\mv{\beta}, \alpha|\kappa,\mv{\rho},\mv{Y})$
w.r.t.\ $\mv{X},\mv{\beta}$ and $\alpha$ (for computational details see Appendix
\ref{app:C1}) and $\mv{\mathcal{I}}$ is the expected Fisher information matrix, (see
Appendix \ref{app:C2}). At each iteration, $\mv{\mathcal{I}}^{-1} \nabla l$ gives a sampling direction from the current state which is similar to a Newton-Raphson step \citep[][Ch. 2]{Newton-Raphson}. Further, the proposal variance,
${\mv{\mathcal{I}}}^{-1}$, accounts for the dependency among the parameters. Due to the GMRF structure of the latent fields, ${\mv{\mathcal{I}}}$ will be a sparse matrix reducing the computations to sampling from a GMRF, for which efficient algorithms exist \citep{rue2004gaussian}.
 
In order to get reasonable acceptance rate, an adaptive MCMC method \citep{AdaptiveMCMC} is used for the step size, $\epsilon$, with the following updating rule;
\begin{equation}
\epsilon_{i+1} = \epsilon_i + \gamma_{i+1} (\widehat{acc}_{\mv{X},\mv{\beta},\alpha}(\epsilon_i)-0.57) 
\label{eq:adaptive_MCMC}
\end{equation}
where $\epsilon_i$ is the step size for the $i^\text{th}$ MCMC iteration, $\gamma_{i} = i^{-1/2}$, $\widehat{acc}$ is the acceptance probability of the $i$th step, and $0.57$ is the target acceptance rate for a MALA proposal as suggested in \citet{acc_rate}.

\subsection{Updating $\kappa$ and $\mv{\rho}$}
The second block is updated using a combination of the conjugate posterior for $[\mv{\rho}|\mv{X},\kappa]$ and a Metropolis-Hastings random walk (in log scale) for $[\kappa|\mv{X}]$. 
The joint posterior of $[\kappa, \mv{\rho}|\mv{X}]$ can be written as
\begin{equation}
\mathbb{P}(\kappa, \mv{\rho}|\mv{X}) = \mathbb{P}(\mv{\rho}|\mv{X},\kappa)\cdot \mathbb{P}(\kappa|\mv{X}).
\label{eq:post-kappa-rho}
\end{equation}
Due to the conjugate prior for $\mv{\rho}$ the conditional posterior for $\mv{\rho}$ is inverse Wishart; 
\begin{equation}
[\mv{\rho}|\kappa, \mv{X}] \propto IW(a_\rho\mathbb{I} + \mv{x}^\trsp \mv{Q}(\kappa) \mv{x},N+b_\rho)
\label{eq:wishart}
\end{equation}
with $\mv{x}$ being a $N\times d$ matrix given by $\mv{x}= [X_1,\cdots, X_d]$. 
The conjugacy makes it possible to marginalize over $\mv{\rho}$ (see Appendix \ref{app:integral}) giving
\begin{equation}
\begin{split}
\mathbb{P}(\kappa|\mv{X}) & \propto \int \mathbb{P}(\mv{\rho}|\kappa, \mv{X})\mathbb{P}(\kappa) d\mv{\rho} \propto \dfrac{a_\rho^{\frac{db_\rho}{2}}|\mv{Q}(\kappa)|^{\frac{d}{2}}}{|a_\rho\mathbb{I} + \mv{x}^\trsp \mv{Q}(\kappa) \mv{x}|^{\frac{N+b_\rho}{2}}}\mathbb{P}(\kappa).
\end{split}
\label{eq:kappa}
\end{equation}

Samples from $[\kappa, \mv{\rho}|\mv{X}]$ are now obtained by first sampling from the posterior \eqref{eq:kappa} using a Metropolis-Hastings step random-walk proposal in log scale, 
\begin{align*}
  \log \kappa^* &= \log \kappa + \epsilon_\kappa, & \epsilon_\kappa &\sim \mathcal{N}(0,\sigma_\kappa^2).
\end{align*}
Given a proposal $\kappa^*$, $\mv{\rho}^*$ is sampled from
\eqref{eq:wishart}. These two steps can be seen as a joint 
Metropolis-Hastings step for $\mv{\rho}$ and $\kappa$ with proposal density
$q(\kappa^*,\mv{\rho}^*|\kappa,,\mv{\rho}) =
\mathbb{P}(\mv{\rho}^*|\mv{X},\kappa^*) \cdot q(\kappa^*|\kappa)$ and acceptance ratio: 
\begin{equation}
  \begin{split}
    \text{acc}_{\kappa,\mv{\rho}} &= \min \left( 1 , 
    \dfrac{\mathbb{P}(\kappa^*,\mv{\rho}^*|\mv{X})}{
      \mathbb{P}(\kappa, \mv{\rho}|\mv{X})} \cdot
    \dfrac{q(\kappa, \mv{\rho} |\kappa^*, \mv{\rho}^*)}{
      q(\kappa^*, \mv{\rho}^*|\kappa, \mv{\rho})} \right) 
  \\ &=
  \min \left( 1 , 
    \dfrac{\mathbb{P}(\mv{\rho}^*|\kappa^*,\mv{X}) \cdot \mathbb{P}(\kappa^*|\mv{X})}{
      \mathbb{P}(\mv{\rho}|\kappa, \mv{X}) \cdot \mathbb{P}(\kappa|\mv{X})} 
    \cdot
    \dfrac{\mathbb{P}(\mv{\rho}|\mv{X},\kappa) \cdot q(\kappa|\kappa^*)}{
      \mathbb{P}(\mv{\rho}^*|\mv{X},\kappa^*) \cdot q(\kappa^*|\kappa)} \right) 
  \\ &= 
  \min \left( 1, 
    \dfrac{\mathbb{P}(\kappa^*|\mv{X})}{\mathbb{P}(\kappa|\mv{X})} \cdot
    \dfrac{\kappa^*}{\kappa} \right).
  \end{split}
  \label{eq:acc-a-rho}
\end{equation}
Since the acceptance ratio depends only on $\kappa$ we can delay the sampling of
$\mv{\rho}^*$ until we know if the suggested $\kappa^*$ has been accepted. 

The proposal variance, $\sigma^2_\kappa$, is determined using an adaptive scheme
similar to \eqref{eq:adaptive_MCMC}, with target acceptance rate of $0.4$
\citep{RoberGG1997_7}. The difference in target acceptance rate is due to the
difference between MALA and random-walk Metropolis-Hastings \citep[see][for a
discussion]{Rosen2011_Brooks}.

\section{Uncertainty}
\label{subsec:Uncertainty}
\begin{figure}[H]
\noindent\makebox[\textwidth]{
\includegraphics[scale=0.65]{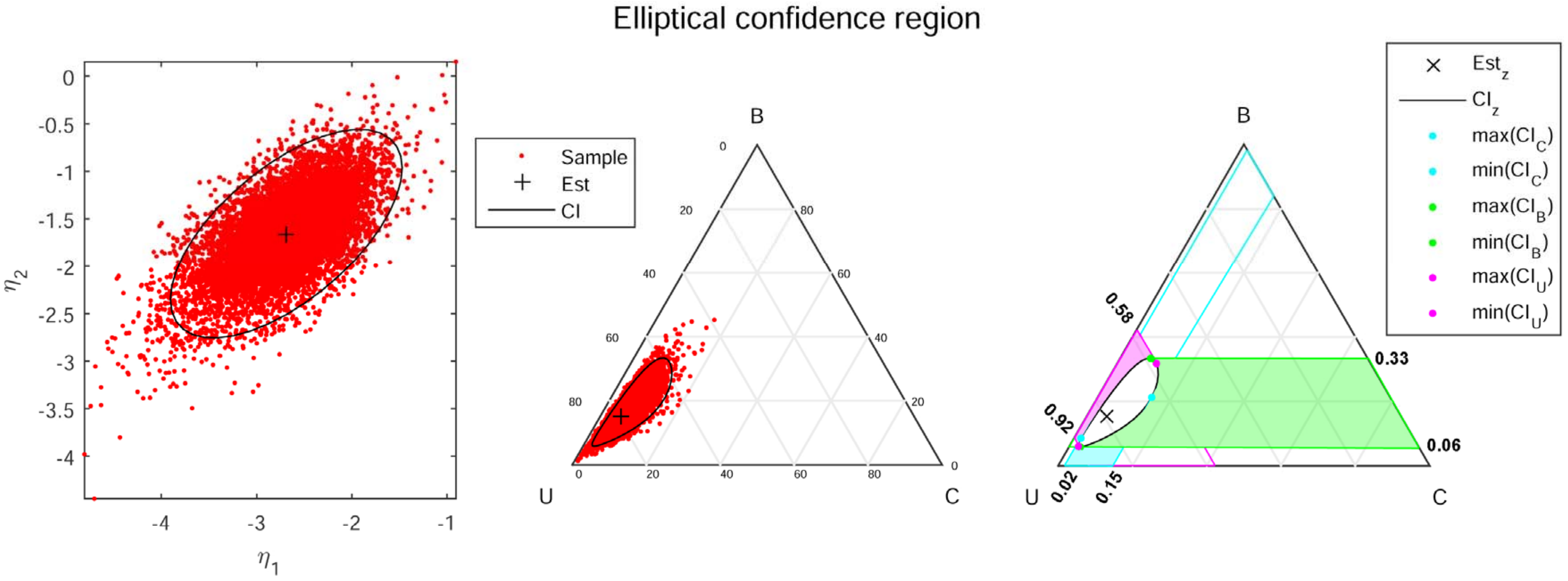}}
\caption{The left plot shows the $95\%$ elliptical confidence region for the $\mv{\eta}$ samples at location $s$. The middle ternary diagram shows the transformed samples and ellipse. The right hand ternary diagram shows the joint maximum and minimum in each composition, C, B, and U; together with the confidence interval for the other two compositions.}
\label{fig:ellipse}
\end{figure}

To obtain uncertainties in the composition estimates at each location, we use the MCMC samples of $\mv{\eta}$ at each location. Given the model structure with a Gaussian prior for $\mv{\eta}$ we base the joint confidence regions for the composition estimates on the elliptical confidence regions obtained for multivariate Gaussian distributions. Using the sample mean, $\mv{\mu}$, and the sample covariance, $\mv{\Sigma}$, in the MCMC samples, we construct the confidence region for each location as the ellipse 
\begin{equation}
  (\mv{\eta}-\mv{\mu})^\trsp\mv{\Sigma}^{-1}(\mv{\eta}-\mv{\mu}) = C_\alpha.
\label{eq:ellipse}
\end{equation}
The quantile $C_\alpha$ is taken as the $\alpha$-quantile of the above squared Mahalanobis distance computed for all the MCMC samples (for a multivariate Gaussian $C_\alpha=\chi^2_\alpha(d)$). Thereafter, the confidence ellipse is transformed from $R^d$ to $(0,1)^D$ using \eqref{eq:link}. The new ternary region is considered as $95\%$ confidence region for the transformed $\mv{\eta}$, i.e.\ the composition estimates. The procedure is illustrated for $D = 3$ in Fig. \ref{fig:ellipse}.

To illustrate the changes in compositions, we choose the maximum and minimum along each dimensions of the ternary plot, i.e. in each component. This way, we get a joint lower bound (minimum) and upper bound (maximum) for each composition together with the corresponding changes in the other compositions ``most likely'' to occur at the bounds (Fig. \ref{fig:ellipse}).

In addition, we compute prediction regions for the compositions by simulating new Dirichlet observations for each MCMC sample of $\mv{\eta}$ and $\alpha$. These D-composition Dirichlet-simulations are then transformed to $\mathbb{R}^d$ using the link function. The procedure above is then used to obtain prediction ellipses and ternary prediction regions.

\section{Application}
\label{sec:application}
The model presented in Sec. \ref{sec:model} was applied to the Pollen data with the goal of reconstructing past land cover over Europe. Two versions of the model in \eqref{eq:eta} were considered:~\begin{inparaenum}[1)] \item a full spatial model with $\mv{\eta}_{all} = \mv{B}\mv{\beta} + \mv{X}$, and \item a regression model with no spatial structure where $\mv{\eta}_{all}=\mv{B}\mv{\beta}$; the regression model is included to allow an evaluation of the need for spatial structure\end{inparaenum}. The intrinsic GMRF model ($\kappa=0$), also considered in \citet{Pirzamanbein2014}, performed similar to or slightly worse than the full spatial model with Dirichlet observations, hence those results have been excluded for brevity.

The remainder of this section consists of: a description of the data (Sec. \ref{subsec:data}), parameters estimation and spatial reconstruction results for the models (Sec. \ref{subsec:result}), and validation of model performance (Sec. \ref{subsec:validation}).

\subsection{Data}
\label{subsec:data}
The data used for the reconstruction of past land-cover composition over Europe consists of pollen-based REVEALS estimates (here called pollen-based land-cover composition data --- PbLCC) of the three land cover types: Coniferous forest, Broadleaved forest and Unforested land. The PbLCC data was obtained using the REVEALS model \citep{sugita2007Reveals} and a detailed description of the data is given by \citet{trondman2015}. REVEALS is mechanistic model that takes into account the size of sedimentary basins and inter-taxonomic differences in pollen productivity and dispersal to estimate regional vegetation cover from pollen records. \citeauthor{trondman2015} applied REVEALS to 636 pollen records from lakes and bogs; producing estimates of regional land cover for 25 plant taxa which were then grouped into the 3 land cover types \citep[see Table 1 and Appendix S2 in][for details]{trondman2015}. The regional estimates from REVEALS were obtained for $1^\circ\times 1^\circ$ grid cells, with each estimate being based on the pollen records from all lakes and bogs within that grid cell \citep{Hellman2008,trondman2015}. The $1^\circ\times 1^\circ$ grid is an appropriate scale for climate models, which currently work at this or higher resolutions \citep{trondman2015}. Since the sedimentary pollen records used by REVEALS are obtained from lakes and bogs the grid based REVEALS estimates are limited to providing land cover in grid cells surrounding the lakes and bogs. This leads to a PbLCC dataset with incomplete coverage across Europe, which needs to be interpolated to produce land cover compositions for the entire region.

The PbLCC data are available for five time periods centred around 1900, 1725 and
1425 CE, 1000 and 4000 BCE, with 175, 181, 193, 204 and 196 observed grid cells,
respectively \citep{trondman2015}; strictly the time periods are: present--1850 CE, 1850--1600 CE, 1600--1250 CE, 1250--750 BCE, and 4250--3750 BCE; where ``present'' should be interpreted as the most recent pollen records recovered at each site. These time periods are commonly used in both climate modelling and palaeoecological studies since they represent major climatic and historical events; Recent Past, Little Ice Age, Black Death, Late Bronze Age, and Early Neolithic.

To capture large scale structures in the land-cover composition, covariates consisting of potential natural vegetation cover adjusted for human land use, and elevation were used. The choice of covariates was based on the best model found in \citet{Pirzamanbein2014}, and detailed descriptions of the covariates can be found in that paper. Here we only provide a brief summary.

Dynamic vegetation model based estimates of climate-induced potential natural vegetation, for the study area and specified time periods, were obtained using the LPJ-GUESS model \citep{smith2001representation}. To account for human land use, the potential natural vegetation was adjusted for anthropogenic deforestation using the KK10 scenarios of \citet{kaplan2009prehistoric}. The KK10 scenarios provide assessments of human induced deforestation based on estimates of past human population densities, land area required for food production to sustain that population, and a model of land suitability for food production. Combining the potential natural vegetation cover from LPJ-GUESS and the KK10 scenarios of deforestation resulted in a land cover covariate, denoted LPJ-GUESS$_{\text{KK}}$.

The elevation data were obtained from the Shuttle Radar Topography Mission \citep{elevation}\footnote{downloaded from \url{ftp://topex.ucsd.edu/pub/srtm30_plus/} on 2011–09–03} and upscaled by averaging from the original resolution of 3 arc-seconds to the  $1^\circ\times 1^\circ$ grid cells. The upscaled data was truncated to $\geq 0$ to handle a few grid cells along the Norwegian coast which otherwise would have negative average elevation due to the presence of deep coastal fjords.

Since the potential land cover, LPJ-GUESS$_{\text{KK}}$, is compositional it
was transformed using \eqref{eq:link}, and the covariate matrix, $\mv{B}$
consisted of the following columns: $B_0$ -- intercept; $B_1,B_2$ --
additive log ratio transformed LPJ-GUESS$_{\text{KK}_{1,2}}$; and $B_3$ -- elevation.

To evaluate our results we used present-time European forest maps compiled by
the European Forest Institute. These maps are based on a combination of
satellite data (NOAA-AVHRR) and national forest-inventory statistics from
$1990$--$2005$ \citep{paivinen2001combining, schuck2002compilation}
\footnote{downloaded from the European Forest Institute webpage \url{http://www.efi.int/portal/virtual_library/information_services/mapping_services/forest_map_of_europe}}. 
The European Forest Institute forest maps (EFI-FM; with proportions of coniferous- and broadleaved-forest cover) were upscaled by averaging from $1$ km $\times 1$ km to $1^\circ\times 1^\circ$ resolution. The proportions of unforested area were
calculated by subtracting the total sum of forested cover from $1$.

\subsection{Results}
\label{subsec:result}
To estimate the parameters for each model, we ran $100\,000$ MCMC iterations with a burn-in sample size of $10\,000$. Diagnostics for the chains indicate a fast convergence for $\alpha,\mv{\rho}$ and $\mv{\beta}$; autocorrelation plots show good mixing of all parameters after burn-in.

Parameter estimates for the 1900 CE time period are given in Table \ref{tab:para1900}; the parameter estimations for the other time periods can be found in Appendix \ref{app:para-est}. As expected the $\alpha$ estimate for the regression model is lower than for the full spatial model, indicating higher observational variation in the regression model. Note that not all the regression coefficients, $\mv{\beta}$, are significant. However we cannot exclude a covariate from one component of the composition and keep it for the other components. Similarly for the transformed potential land cover composition we need to include all or none of the compositional components.

\begin{table}[H]
\caption{Parameter estimates (Est) and $95\%$ confidence intervals (CI) for the two models (Full --- spatial model and RM --- regression model) fitted to the PbLCC data from the 1900 CE time period.}
\label{tab:para1900}
\centerline{
\begin{tabular}{crlrl}
\multicolumn{5}{l}{1900 CE}\\
\hline
\multicolumn{1}{c}{}&\multicolumn{2}{c}{Full}&\multicolumn{2}{c}{RM}\\
\cline{2-5}
\multicolumn{1}{c}{Parameter}&\multicolumn{1}{c}{$\text{Est}$}&\multicolumn{1}{c}{(CI)}&\multicolumn{1}{c}{$\text{Est}$}&\multicolumn{1}{c}{(CI)}\\
\hline
$\alpha$	&	10.86	&	(	8.10	,	15.41	)	&	6.36	&	(	5.58	,	7.18	)	\\
$\kappa$	&	0.28	&	(	0.14	,	0.45	)	&	-	&			-			\\
$\rho_{11}$	&	0.78	&	(	0.12	,	2.61	)	&	-	&			-			\\
$\rho_{12}$	&	0.57	&	(	0.05	,	1.96	)	&	-	&			-			\\
$\rho_{22}$	&	0.60	&	(	0.10	,	1.97	)	&	-	&			-			\\
$\beta_{10}$	&	-0.68	&	(	-1.64	,	0.15	)	&	-0.13	&	(	-0.25	,	-0.02	)	\\
$\beta_{11}$	&	0.16	&	(	0.08	,	0.24	)	&	0.24	&	(	0.22	,	0.27	)	\\
$\beta_{12}$	&	0.02	&	(	-0.09	,	0.14	)	&	-0.03	&	(	-0.09	,	0.02	)	\\
$\beta_{13}$	&	0.05	&	(	-0.15	,	0.26	)	&	-0.10	&	(	-0.19	,	-0.01	)	\\
$\beta_{20}$	&	-0.94	&	(	-1.83	,	-0.22	)	&	-0.38	&	(	-0.51	,	-0.26	)	\\
$\beta_{21}$	&	0.04	&	(	-0.05	,	0.12	)	&	0.13	&	(	0.11	,	0.16	)	\\
$\beta_{22}$	&	0.01	&	(	-0.09	,	0.11	)	&	-0.05	&	(	-0.10	,	0.00	)	\\
$\beta_{23}$	&	-0.04	&	(	-0.24	,	0.16	)	&	-0.24	&	(	-0.34	,	-0.14	)	\\
\hline
\end{tabular}}
\end{table}

Reconstructions of the land-cover composition for the two models and the 1900 CE time period are shown in Fig. \ref{fig:full}. Results for the other time periods are available in Appendix \ref{app:results}. Figure \ref{fig:full} shows that the land-cover reconstructions from the two models captured the structure in the PbLCC data. However, the results from regression model is smoother than from the Full model. The Full model better captures the high abundance of unforested land in Poland, Denmark and south east Norway.

The uncertainties in the land-cover reconstructions were computed using the method described in Sec. \ref{subsec:Uncertainty}. Results for the $1900$ CE time period are presented in Fig.~\ref{fig:PR_CR_AD1900} and \ref{fig:results_CI_AD1900}, with results for the remaining time periods given in Appendix \ref{app:results_CI}. The confidence and prediction regions represent the uncertainty in the latent field reconstruction, $\mv{z_s}$ and the potential uncertainty in new PbLCC data, $\mv{y_s}$, for a given grid cell, respectively. In general the full spatial model has larger confidence regions but smaller prediction regions than the regression model (Fig.~\ref{fig:PR_CR_AD1900}). This is due to the spatial component in the Full model being able to better capture spatial variation resulting in a lower uncertainty (larger $\alpha$) in the Dirichlet observations as compared to regression model. The maps of confidence regions (Fig.~\ref{fig:results_CI_AD1900}) illustrate rather large uncertainties in the predicted land-cover composition in general, and especially for Southeast Europe, a region with very few observations.

\begin{figure}[H]
\noindent\makebox[\textwidth]{
\includegraphics[scale=0.55]{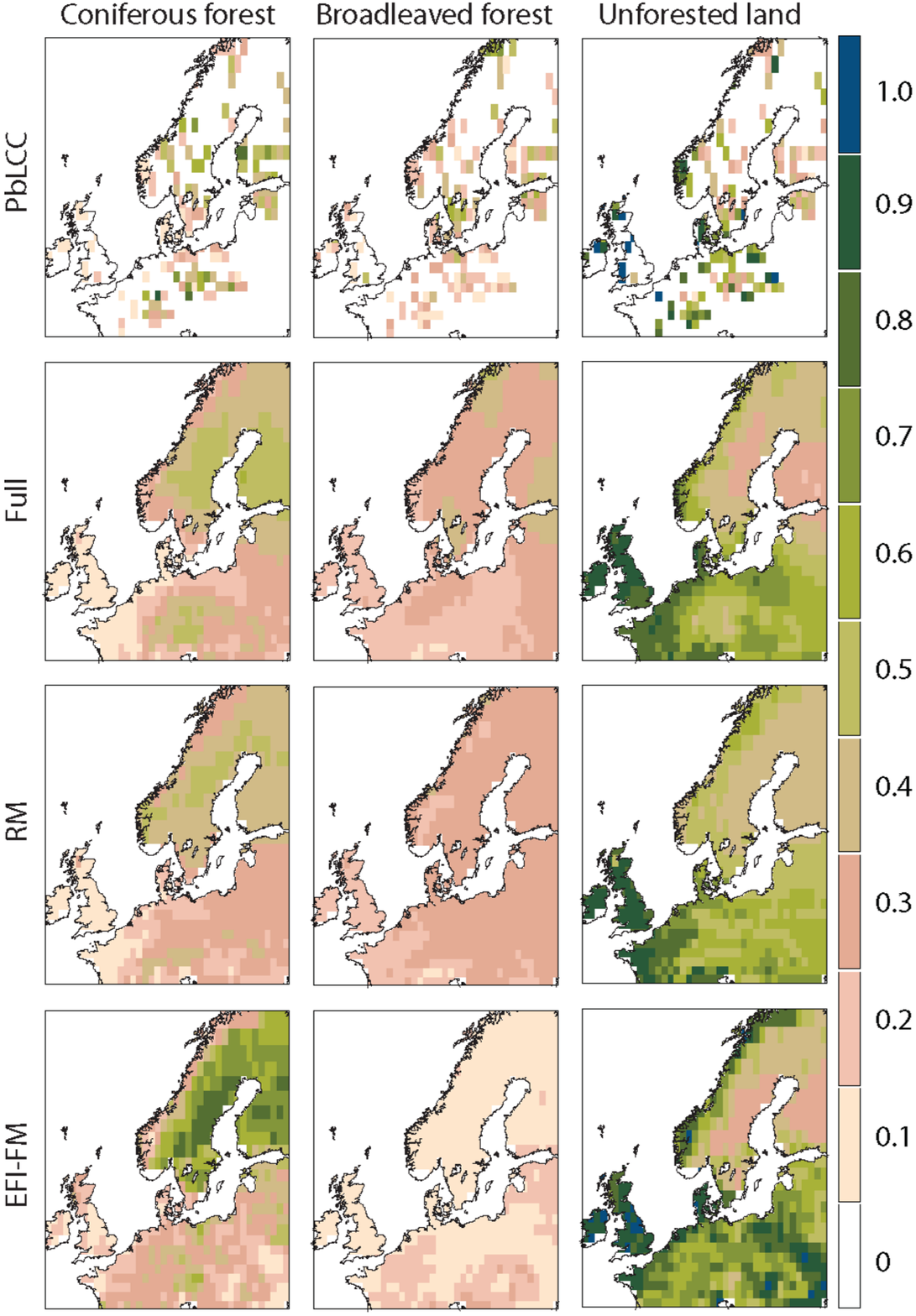}}
\caption{Results for the 1900 CE time period: the top row shows the PbLCC data from REVEALS, the bottom row shows the EFI-FM and the remaining rows show the reconstructions for the full spatial model (Full) and the regression model (RM).}
\label{fig:full}
\end{figure}

\begin{figure}[H]
\noindent\makebox[\textwidth]{
\includegraphics[scale=0.7]{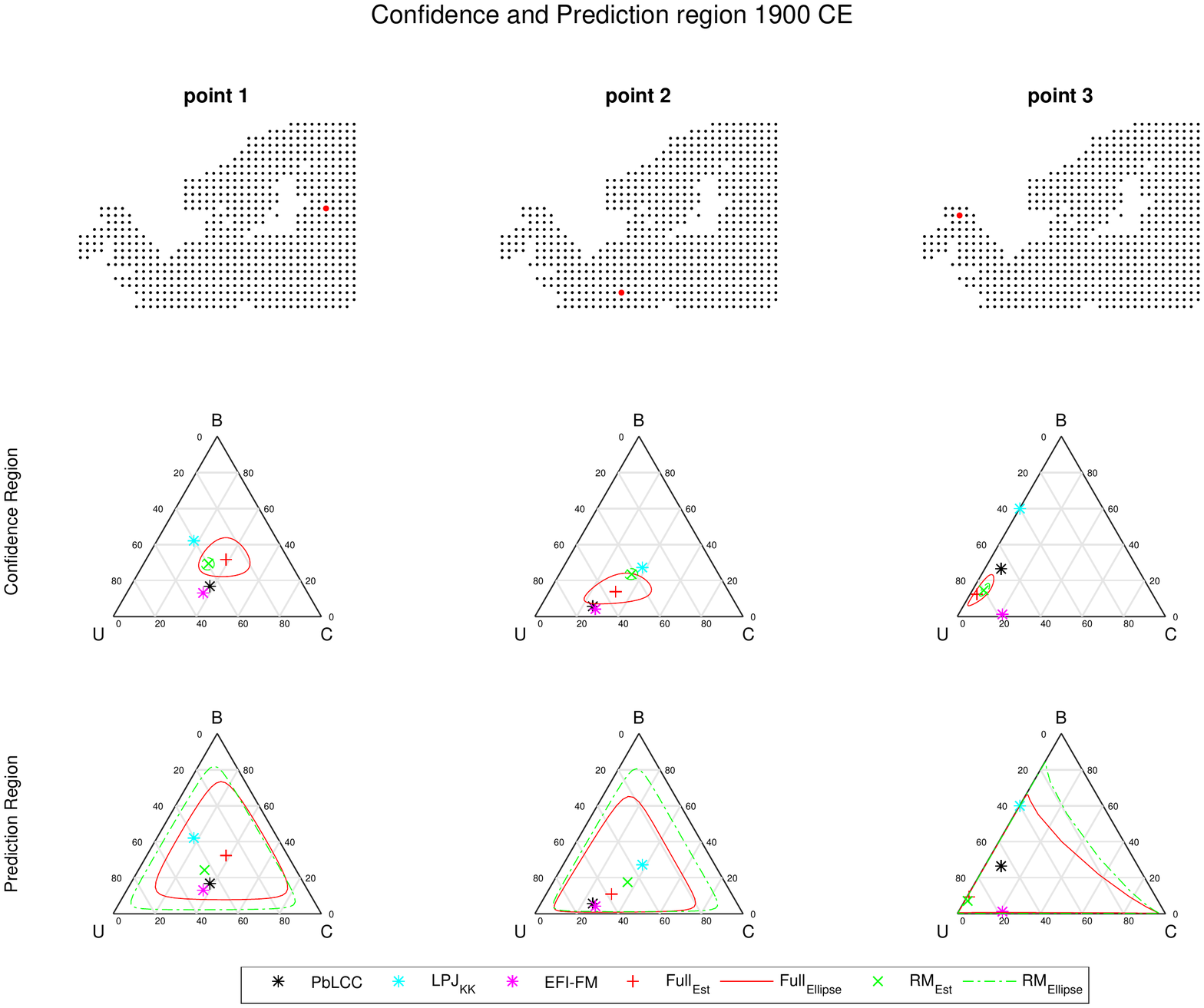}}
\caption{The first row shows the locations of the three selected grid cells. The second row shows the ternary confidence regions and the land-cover reconstructions for the two models (Full --- spatial model and RM --- regression model) together with the PbLCC data from REVEALS, the LPJ-GUESS$_{\text{KK}}$ land cover covariate and the EFI-FM for each location. The third row shows the ternary prediction regions.}
\label{fig:PR_CR_AD1900}
\end{figure}

\begin{figure}[H]
\noindent\makebox[\textwidth]{
\includegraphics[scale=0.65]{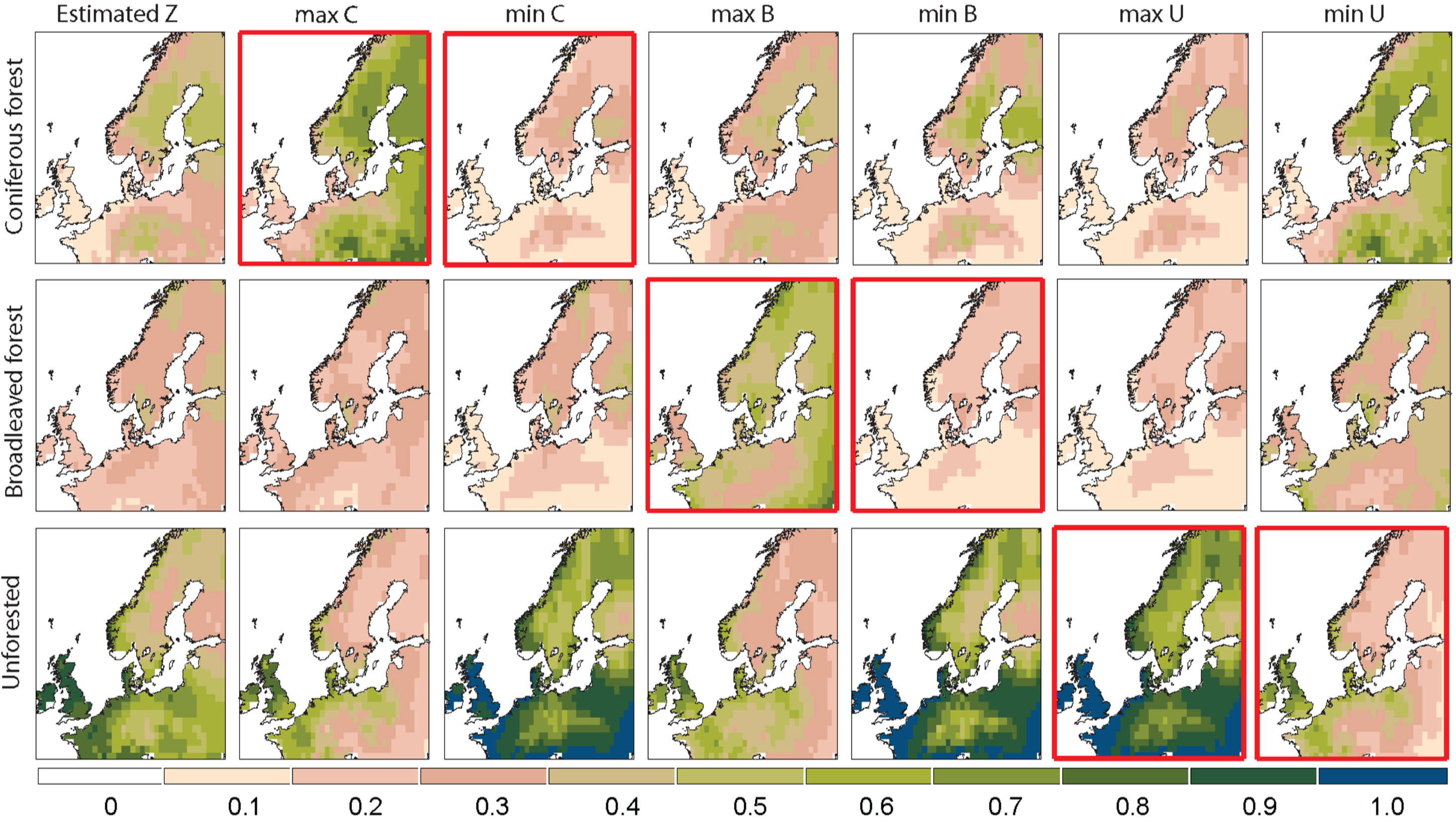}}
\caption{The first column shows the reconstructed land-cover composition for the 1900 CE time period using the full spatial model. Columns 2 and 3, row 1 (with thick/red axes), show the maximum and minimum of $95\%$ elliptical confidence regions for Coniferous; rows 2 and 3 give the corresponding Broadleaved and Unforested compositions. Columns 4 and 5 (row 2 with thick/red axes) gives the bounds for the Broadleaved composition while columns 6 and 7 show the bounds for Unforested land (row 3 with thick/red axes). The concept of joint confidence interval for compositions is shown in Fig \ref{fig:ellipse}.}
\label{fig:results_CI_AD1900}
\end{figure}

\subsection{Validation}
\label{subsec:validation}
To evaluate the performance of the models, we compared the land-cover reconstructions for 1900 CE to the EFI-FM by computing the average compositional distances (ACD). The compositional distances \citep{aitchison2000logratio, Aitch1992_24, aitchison1986statistical} were computed for each location as
\begin{equation}
\text{ACD}(\mv{u},\mv{v})=[(\mv{u}-\mv{v})^T\mv{J}^{-1}(\mv{u}-\mv{v})]^{1/2}
\label{eq:distance}
\end{equation}
where $\mv{u}$ and $\mv{v}$ are additive log-ratio transforms of the
compositions to be compared and $\mv{J}$ is a $d \times d$-matrix with elements
$J_{p,l}=2$~ if~ $p=l$, and $J_{p,l}=1$~ if~ $p\neq l$. These compositional
distances are then averaged over all grid cells. 
In terms of the original compositions, $\mv{p^u}$ and $\mv{p^v}$, the distance
in \eqref{eq:distance} can be written as \citep{aitchison2000logratio}
\begin{align*}
\text{ACD}(\mv{p^u},\mv{p^v}) &= \left[
  \sum_{i=1}^D \left( \log \frac{p^u_i}{g(\mv{p^u})} - 
    \log \frac{p^v_i}{g(\mv{p^v})}
  \right) \right]^{1/2},
\end{align*}
where $g(\mv{p})$ is the geometric mean, 
$g(\mv{p}) = \sqrt[D]{ p_1 p_2 \cdots p_D }$.

Although a temporal misalignment exists between the PbLCC data (PbLCC data are from 1850 to the present) and the EFI-FM (inventory and satellite data are from 1990-2005); EFI-FM provides the best complete and consistent land cover map of Europe for present times, making it a reasonable choice for the comparison. Figure \ref{fig:full} shows the maps of PbLCC data and EFI-FM. The main differences between the EFI-FM data and the PbLCC data for the 1900 CE time period are:~\begin{inparaenum}[1)]\item a lower abundance of broadleaved forest around most of Europe, \item a higher abundance of coniferous forest in Sweden and Finland, and \item a higher abundance of unforested land in North Norway\end{inparaenum}\ in the EFI-FM data than in the PbLCC data. Compositional distances between land-cover reconstructions and EFI-FM were computed using \eqref{eq:distance} and averaged over all grid cells. The resulting ACD are $1.4757$ and $1.5025$ for the full spatial model and the regression model, respectively. This indicates that the full spatial model provides a reconstruction closer to EFI-FM than the regression model.

These results can also be compared to a model with Gaussian observations of
transformed latent fields, \citep{Pirzamanbein2014}. The resulting ACD of the
Gaussian observation models compare to the EFI-FM are $1.6007$ (for the intrinsic GMRF model) and $1.6140$ for the regression model. The differences between what
\cite{Pirzamanbein2014} reported ($1.5201$ and $1.5177$, respectively) and our results using their models are due to an increase in available data leading to more grid cells in our reconstructions. These results indicate smaller distances between the land-cover reconstructions and EFI-FM for the models with Dirichlet observations proposed in this paper compared to similar models with Gaussian observations.

Since no ground truth exists for the other time periods, we applied a 6-fold cross-validation scheme for the models for each of the five time periods \citep[][Ch. 7.10]{crossvalidation}. The cross-validation was run for 10 different, randomly selected 6 folds to assess the  variability due to different cross validation groupings. Average compositional errors and standard deviations are shown in Table \ref{tab:cv}. The full spatial model gives the best predictions for all the five time periods.
\begin{table}[h!]
\caption{Average compositional error (and standard deviation) from 10 different 6-fold cross-validations for each of the models, and time periods.}
\label{tab:cv}
\centerline{
\begin{tabular}{l|rl|rl}
\hline
\multicolumn{1}{c}{\multirow{2}{*}{} }& \multicolumn{2}{c}{Full}&\multicolumn{2}{c}{Regression}\\
\multicolumn{1}{c}{Time}& \multicolumn{2}{c}{$\text{CV}_\text{error}$ (sd)} & 
\multicolumn{2}{c}{$\text{CV}_\text{error}$ (sd)}\\
\hline
1900 CE & \textbf{1.0169} & (0.0122)  & 1.1439 & (0.0061)
\\
1700 CE  & \textbf{1.1448} & (0.0084) &	1.2891 & (0.0054)
\\
1400 CE& \textbf{1.2009} & (0.0071) &	1.4061 & (0.0042)
\\
1000 BCE& \textbf{1.3260} & (0.0083)	&	1.5287 & (0.0062)
\\
4000 BCE& \textbf{1.2131} & (0.0109)	&	1.3396 & (0.0045)
\\
\hline
\end{tabular}}
\end{table}

\section{Conclusion}
\label{sec:conclusion}
In this paper we have introduced a model for spatial interpolation of
compositional data that relies on Dirichlet observations of an underlying
multivariate GMRF. In theory the formulation allows for a wide class of
link-functions between the GMRF and the compositional probabilities in the
Dirichlet observations; we used the additive log ratio transformation throughout the paper. Since the sparse structure in the precision matrix of the GMRF carries
over to the expected Fisher information used in MALA, the model formulation with
a latent GMRF allows for fast MCMC-based estimation of parameters and latent
field. As a result our MCMC produced $10$ samples per second using 
{\sc matlab}\textregistered{} on a standard desktop (Intel\textregistered{} 
Core\textsuperscript{TM} $i7-2600$ CPU (2011) with 8 GB memory) for a latent
field with $2160$ nodes (bivariate field on a $27$-by-$40$ grid); resulting in a total run time of less than 3 hours for a chain with $100\,000$ samples.

To evaluate prediction uncertainties we also proposed a method for construction
of joint confidence and prediction regions for the compositions at each
location. The idea behind the method is to use the MCMC samples to first construct elliptical confidence regions for the transformed latent fields; these are then transformed from $R^d$ to $(0,1)^D$ using the inverse link-function, giving confidence regions in compositional space. Having joint confidence regions
for the compositions allowed us to evaluate the behaviour of all components
as each individual component attains their lower and upper bounds in the
confidence regions.

The statistical model was used to reconstruct past land-cover composition over Europe for five time periods using PbLCC data \citep{trondman2015} obtained from the REVEALS model \citep{sugita2007Reveals}. The land-cover reconstructions for the most recent time period were evaluated against present-time forest maps, and reconstructions for all time-periods were evaluated using cross-validation. The evaluations showed that a model containing both explanatory covariates and spatial dependence structure outperformed a model with only covariates, indicating that the addition of a spatial random effect improves predictions. Evaluations using the present-time forest maps showed that a model with Dirichlet observations outperformed previously developed models using Gaussian observations of transformed fields \citep{Pirzamanbein2014}.

\begin{figure}[H]
\noindent\makebox[\textwidth]{
\includegraphics[scale=0.55]{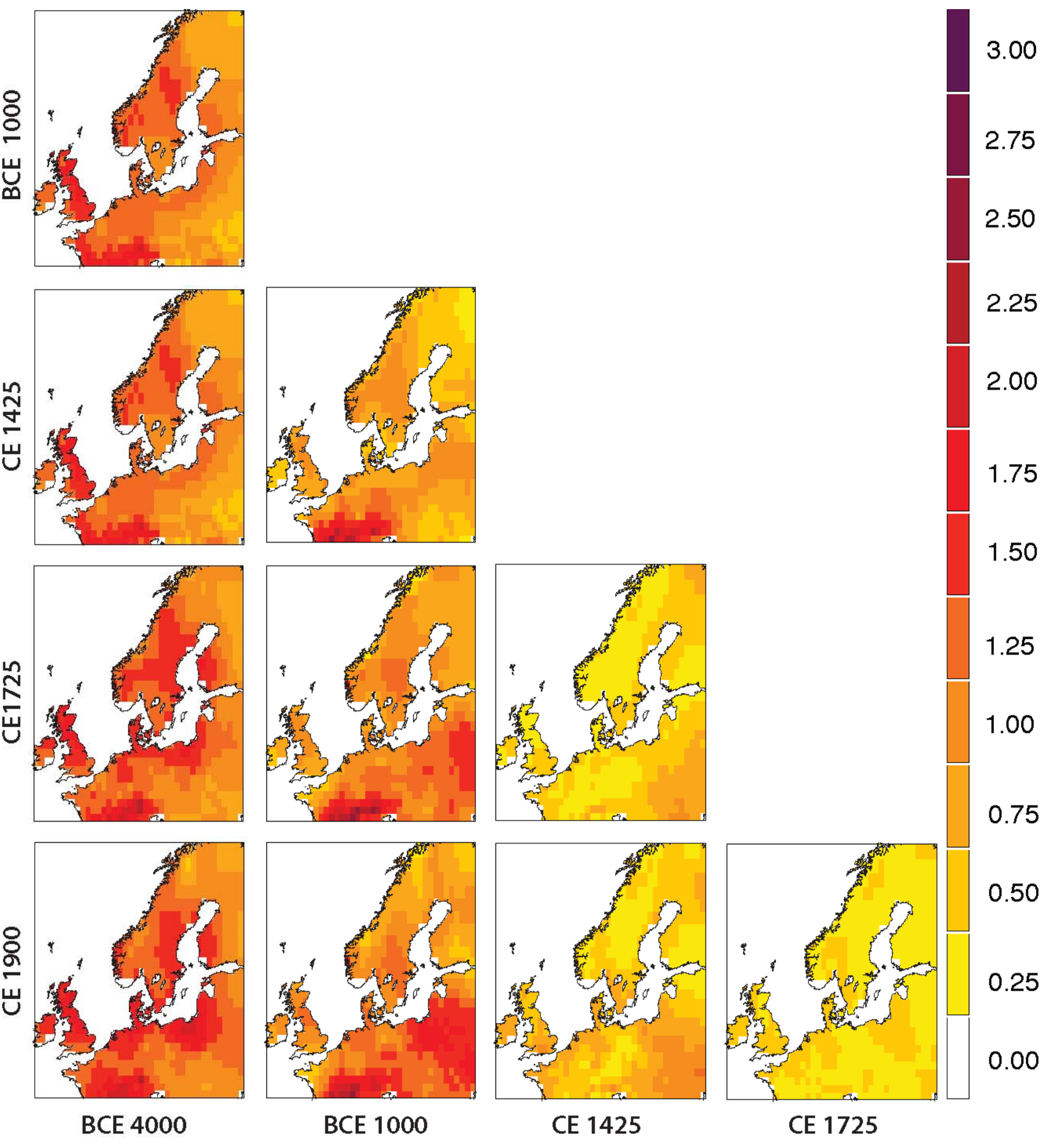}}
\caption{Compositional distances between the Full model land-cover reconstructions for the different time periods.}
\label{fig:time_diff_full}
\end{figure}

The reconstructed maps of land-cover composition can be used both in studies of climate models and to analyse changes in land-cover composition during the past millennia. For example, Fig.~\ref{fig:time_diff_full} uses the compositional distances \eqref{eq:distance} to illustrate the changes in land-cover composition between the five time periods considered in this study. 
This simple analysis shows that the largest changes in land cover between 4000 BCE and 1900 CE have occurred in Switzerland and Central France; along the North Sea coast in the UK, the Low Countries, Denmark, and southern Norway; and along the south Baltic coast in northern Germany and Poland.

The reconstructions of past land-cover composition obtained here are encouraging, as they clearly show the ability to recover continuous maps of past land cover from PbLCC data. The reconstructions from the full spatial model appear to conserve the information and trends from the pollen-based REVEALS estimates of past land cover \citep[as discussed in][]{trondman2015} the best. They are also clearly better than previous spatial reconstructions in terms of e.g.\ the degree of openness and tree cover in the 
northernmost parts of Europe and the western coasts of Norway. Our future goal is 
to use these land-cover reconstructions in climate modelling studies and to gain insight into the effect of past anthropogenic deforestation. It is outside the scope of this
paper to provide a discussion of the land-cover reconstructions in terms of historical changes in vegetation abundance, land cover and human impact over the past $6\,000$ years. However, the methods developed here provide (some of) the tools needed for such a discussion. 

\bibliographystyle{abbrvnat}
\bibliography{Journals_abrv,second_bib}

\newpage
\appendix
\section{Derivatives and Fisher Information of $[\mv{\eta}_{all}, \alpha|\mv{Y}]$}
\label{app:d_FI}
To construct the MALA updates for $\mv{\eta}_{all} = \mv{B}\mv{\beta}+ \mv{X}, \alpha$ we need the derivatives
and Fisher information of the log-posterior \eqref{eq:post_eta_alpha},
\begin{equation}
  \begin{split}
    l(\mv{X},\mv{\beta},\alpha|\mv{Y}) =& 
    \log \left(\prod_{s=1}^{N_o}
      \mathbb{P}(\mv{y}_s|f_s\big(\mv{A}\mv{\eta}_{all}\big),\alpha)
      \mathbb{P}(\mv{X}|\kappa,\mv{\rho})
      \mathbb{P}(\mv{\beta})
      \mathbb{P}(\alpha) \right)
    \\ =&
    \sum_{s=1}^{N_o} \log \Gamma(\alpha)
    - \sum_{s=1}^{N_o} \sum_{k=1}^D \log \Gamma(\alpha z_{s,k})
    \\ &
    + \sum_{s=1}^{N_o}\sum_{k=1}^D (\alpha z_{s,k}-1)\log y_{s,k}
    - \frac{1}{2} \mv{X}^{\trsp} \left( 
      \mv{\rho}^{-1}\otimes \mv{Q}(\kappa) \right) \mv{X} 
    \\ &
    - \frac{q_{\mv{\beta}}}{2} \mv{\beta}^{\trsp} \mv{\beta}
    + (a_\alpha-1) \log(\alpha) - \alpha b_\alpha
    + \text{const}.
  \end{split}
  \label{eq:app_log_post_eta_alpha}
\end{equation}
Here $\mv{A}\mv{\eta}_{all}$ is the latent $\R^d$-field at observed locations, $\{u_s\}_{s=1}^{N_o}$,
$\mv{z}_s$ is the corresponding $D$-composition (i.e.\ defined on
$(0,1)^D$, with $d=D-1$), and $\text{const}$ is an additive constant.
Before computing derivatives of the log-posterior, $l(\mv{X},\mv{\beta},
\alpha|\mv{Y})$, we need some results for the compositional transformation.

\subsection{Derivatives of Compositional Transforms}
The compositional transform used in this paper is the additive log-ratio,
\eqref{eq:alr}, with inverse
\begin{equation}
  z_k =
  \begin{cases}
    \frac{\exp(\eta_k)}{1+\sum_k^{d} \exp(\eta_k)}, & \text{if}\ k=1,\ldots,D-1
    \\
    \frac{1}{1+\sum_k^{d} \exp(\eta_k)}, & \text{if}\ k=D.
  \end{cases}
  \label{eq:invalr}
\end{equation}
Here $z$ is a $D$-compositional value (i.e.\ $(0,1)^D$) and $\eta$ is
$\R^{D-1}$.

For the MALA-computations the first and second derivatives of the inverse
transformation are needed. These can be expressed in terms of the compositions,
$z_k$; for the first derivatives
\begin{equation}
  \frac{\partial z_k}{\partial \eta_i} =
  \begin{cases}
    z_k(1-z_k) & \text{if}\ k = i,
    \\
    -z_k z_i & \text{if}\ k \neq i.
  \end{cases}
  \label{eq:dalr}
\end{equation} 
and for the second derivatives
\begin{equation}
  \begin{split}
    \frac{\partial^2 z_k}{\partial \eta_i \partial \eta_j} &= 
    \begin{cases}
      z_k (1-z_k) (1-2z_k), & \text{if}\ i=j, k=i
      \\
      -z_k z_i (1-2z_i), & \text{if}\ i=j, k \neq i
      \\
      -z_j z_k (1-2z_k), & \text{if}\ i \neq j, k=i 
      \\
      2 z_k z_i z_j, & \text{if}\ i \neq j, k \neq i, k \neq j, 
    \end{cases}
  \end{split}
  \label{eq:d2alr}
\end{equation}
the case $i \neq j, k=j$ is obtained by symmetry.

One consequence of the sum to one constraint of compositional data is that the
derivatives \eqref{eq:dalr} and second derivatives \eqref{eq:d2alr} sum to zero:
\begin{equation}
  \begin{split}
    \sum_{k=1}^D \frac{\partial z_k}{\partial \eta_i} 
    &=
    z_i(1-z_i)  - \sum_{k \neq i} z_iz_k
    =
    z_i \left( 1 - \sum_{k=1}^D z_k \right) = 0
    \\
    \sum_{k=1}^D \frac{\partial^2 z_k}{\partial \eta_i^2} 
    &=
    z_i (1-z_i) (1-2z_i) - \sum_{k \neq i} z_k z_i (1-2z_i)
    = 0
    \\
    \sum_{k=1}^D \frac{\partial^2 z_k}{\partial \eta_i \partial \eta_j} 
    &=
    -z_j z_i (1-2z_i) -z_i z_j (1-2z_j) + \sum_{k \neq i,j} 2 z_k z_i z_j
    = 0
  \end{split}
  \label{eq:appB_sum}
\end{equation}

\subsection{Derivative of $l(\mv{X},\mv{\beta}, \alpha|\mv{Y})$}
\label{app:C1}
Recall that the latent field $\mv{\eta}$ is a linear combination of the mean
zero spatial field(s) $\mv{X}$ and the regression coefficients $\mv{\beta}$
given as
\begin{equation*}
  \mv{\eta} = 
  \mv{A}\big(\bmat{ \mathbb{I} & \mv{B} }
    \bmat{ \mv{X}\\\mv{\beta} }\big) = \bmat{ \mv{A} & \mv{A}\mv{B} }
    \bmat{ \mv{X}\\\mv{\beta} }.
\end{equation*}
Therefore, the updates of $\mv{\eta}$ are done by updating the underlying fields
and regression coefficients. Thus we need the derivatives of $l(\mv{X},\mv{\beta},\alpha|\mv{Y})$, i.e. $\nabla l$, w.r.t\ $\mv{\theta} = \bmat{ \mv{X}^\trsp & \mv{\beta}^\trsp }^\trsp$ and $\alpha$.
\begin{subequations}
  \label{eq:app_dl}
  \begin{align}
  	\nabla_{\mv{\theta}} l(\mv{X},\mv{\beta}, \alpha|\mv{Y}) =&
    \bmat{ \mv{A} & \mv{A}\mv{B} }^\trsp
    \nabla_{\mv{\eta}} \log \mathbb{P}(\mv{Y}|f(\mv{\eta}), \alpha)
    - \bmat{ 
      \left(\mv{\rho}^{-1}\otimes \mv{Q}(\kappa) \right) \mv{X}\\
      q_{\mv{\beta}} \mv{\beta}
     }
    \label{eq:app_dl_theta}
    \\
    \begin{split}
    \frac{\partial l(\mv{X},\mv{\beta}, \alpha|\mv{y})}{\partial \alpha} =&
    \sum_{s=1}^{N_o} \psi(\alpha)
    - \sum_{s=1}^{N_o} \sum_{k=1}^D z_{s,k} \psi(\alpha z_{s,k})
    \\ &
    + \sum_{s=1}^{N_o}\sum_{k=1}^D z_{s,k} \log y_{s,k}
    + \frac{a_\alpha-1}{\alpha} - b_\alpha
    \end{split}
    \label{eq:app_dl_alpha}
  \end{align}
\end{subequations}
where $\nabla_{\mv{\theta}}l$ is the gradient w.r.t.\ $\mv{\theta}$ (a
$Nd$-column vector) and $\psi(\cdot)$ is the digamma function. The elements of the gradient $\nabla_{\mv{\eta}} \log \mathbb{P}(\mv{Y}|f(\mv{\eta}), \alpha)$ 
(a $N_{o}d$-column vector) are
\begin{align*}
    \frac{\partial \log \mathbb{P}(\mv{Y}|f(\mv{\eta}), \alpha)}{\partial \eta_{s,k}} =&
    \sum_{l=1}^D \Bigl( -\alpha \psi(\alpha z_{s,l})
      + \alpha \log y_{s,l} \Bigr)
    \frac{\partial z_{s,l}}{\partial \eta_{s,k}},
\end{align*}
where the derivatives, $\partial z_{s,l}/\partial \eta_{s,k}$, depend on the choice of link function (see \eqref{eq:dalr} for the additive log ratio case).

\subsection{The Fisher Information}
\label{app:C2}
The Fisher information used in the MALA updates is computed as the expectation
of the Hessian over observations, $\mv{Y}$, given all parameters and latent fields:
\begin{equation}
  \mv{\mathcal{I}} = -\pE_{\mv{Y}}\!\left( \mv{H} (\l )
    \,\middle|\, \mv{X}, \mv{\beta}, \alpha, \mv{\rho}, \kappa \right)
  = \bmat{  
    \mv{\mathcal{I}}_{\mv{\theta},\mv{\theta}} & \mv{\mathcal{I}}_{\mv{\theta},\alpha} \\
    \mv{\mathcal{I}}_{\alpha,\mv{\theta}} & \mv{\mathcal{I}}_{\alpha,\alpha}
   }
  \label{eq:app_FI}
\end{equation}
where $\mv{H}(\l)$ is Hessian of $l(\mv{X}, \mv{\beta}, \alpha|\mv{Y})$.
The resulting matrix consists of four blocks: two with second derivatives
w.r.t.\ $\mv{\theta}$ and $\alpha$, and two with cross partial derivatives; each of the blocks is described below. For brevity we use
$\pE\!\left( \mv{H}(l) \,\middle|\, \bullet \right)$ to denote the conditional expectation in \eqref{eq:app_FI}, and note that
\begin{equation}
  \pE\!\left( \log y_{s,k} \,\middle|\, \bullet \right) 
  = 
  \psi\left( \alpha z_{s,k} \right) - 
  \psi\left( \sum_{l=1}^D \alpha z_{s,l} \right)
  =
  \psi(\alpha z_{s,k}) - \psi( \alpha).
  \label{eq:E_logy}
\end{equation}

Similar to \eqref{eq:app_dl_theta} the top left block can be written as
\begin{align*}
  \mv{\mathcal{I}}_{\mv{\theta},\mv{\theta}} =
    \bmat{ \mv{A} & \mv{A}\mv{B} }^\trsp
    \mv{H}_{\mv{\eta}}
    \bmat{ \mv{A} & \mv{A}\mv{B} }
    + \bmat{ 
      \mv{\rho}^{-1}\otimes \mv{Q}(\kappa) & \mv{0} \\
      \mv{0} & q_{\mv{\beta}} \mathbb{I}
     },
\end{align*}
where $\mv{H}_{\mv{\eta}}$ is a symmetric $N_{o}d \times N_{o}d$ matrix with
elements
\begin{align*}
\mv{H}_{\mv{\eta}}^{(sk,s'k')} =
  - \pE\!\left( 
    \frac{\partial^2 \log \mathbb{P}(\mv{Y}|f(\mv{\eta}), \alpha)}{
      \partial \eta_{s,k} \partial \eta_{s',k'}} 
    \,\middle|\, \bullet \right).
\end{align*}
The elements in $\mv{H}_{\mv{\eta}}$ are indexed by their spatial location,
$s=1,\ldots,N_{o}$, and which latent field, $k=1,\ldots,d$, they belong to
(i.e.\ which transformed compositional component).
For elements at different locations 
\begin{equation*}
  \mv{H}_{\mv{\eta}}^{(sk,s'k')} = 0,\quad \text{if}\quad s \neq s',
\end{equation*}
 leaving only 
\begin{align*}
  \mv{H}_{\mv{\eta}}^{(sk,sk')} =&
  - \pE\!\left( 
    \frac{\partial^2 \log \mathbb{P}(\mv{Y}|f(\mv{\eta}), \alpha)}{
      \partial \eta_{s,k} \partial \eta_{s,k'}} 
    \,\middle|\, \bullet \right)
  \\=&
  -  \frac{\partial}{\partial \eta_{s,k'}}
  \pE\!\biggl( 
  \sum_{l=1}^D \Bigl( -\alpha \psi(\alpha z_{s,l})
  + \alpha \log y_{s,l} \Bigr)
  \frac{\partial z_{s,l}}{\partial \eta_{s,k}}
  \,\biggm|\, \bullet \biggr)
  \\=&
  \alpha^2 \sum_{l=1}^D \psi'(\alpha z_{s,l})
  \frac{\partial z_{s,l}}{\partial \eta_{s,k'}}
  \frac{\partial z_{s,l}}{\partial \eta_{s,k}}
  \\&
  + \alpha \sum_{l=1}^D 
  \Bigl(\psi(\alpha z_{s,l}) -
  \pE\!\left(\log y_{s,l} \,\middle|\, \bullet\right)
  \Bigr)
  \frac{\partial^2 z_{s,l}}{\partial \eta_{s,k} \partial \eta_{s,k'}}.
\end{align*}
Using the expectations in \eqref{eq:E_logy} gives
\begin{align*}
  \mv{H}_{\mv{\eta}}^{(sk,sk')} =&
  \alpha^2 \sum_{l=1}^D \psi'(\alpha z_{s,l})
  \frac{\partial z_{s,l}}{\partial \eta_{s,k'}}
  \frac{\partial z_{s,l}}{\partial \eta_{s,k}}
  + \alpha \psi(\alpha) \sum_{l=1}^D 
  \frac{\partial^2 z_{s,l}}{\partial \eta_{s,k} \partial \eta_{s,k'}},
\end{align*}
and with the sum to zero result in \eqref{eq:appB_sum}
the elements of $\mv{H}_{\mv{\eta}}$ simplify to
\begin{align*}
  \mv{H}_{\mv{\eta}}^{(sk,sk')} =&
  \alpha^2 \sum_{l=1}^D \psi'(\alpha z_{s,l})
  \frac{\partial z_{s,l}}{\partial \eta_{s,k'}}
  \frac{\partial z_{s,l}}{\partial \eta_{s,k}}.
\end{align*}

Derivation of \eqref{eq:app_dl_theta} w.r.t.\ $\alpha$ gives
\begin{align*}
  \mv{\mathcal{I}}_{\mv{\theta},\alpha} =&
  - \bmat{ \mv{A} & \mv{A}\mv{B} }^\trsp
  \pE\!\left( \frac{\partial}{\partial \alpha} 
    \nabla_{\mv{\eta}} \log \mathbb{P}(\mv{Y}|f(\mv{\eta}), \alpha)
    \,\middle|\, \bullet \right),
\end{align*}
since only the Dirichlet part of the log-likelihood contributes too the cross-derivatives.
The part concerning the gradient, $\nabla_{\mv{\eta}}
\log \mathbb{P}(\mv{Y}|f(\mv{\eta}), \alpha)$, gives a column vector of length
$N_{o}d$ with elements
\begin{align*}
  \pE\!\left( \frac{\partial^2 \log \mathbb{P}(\mv{Y}|f(\mv{\eta}), \alpha)}{
      \partial\alpha \partial \eta_{s,k}} \,\middle|\, \bullet \right)
  =&
  \sum_{l=1}^D \Bigl( 
  - \psi(\alpha z_{s,l})
  - \alpha z_{s,l} \psi'(\alpha z_{s,l}) +
  \\ & \phantom{\sum_{l=1}^D \Bigl(}
  + \pE\!\left(\log y_{s,l} \,\middle|\, \bullet \right)
  \Bigr) \frac{\partial z_{s,l}}{\partial \eta_{s,k}},
  \\ =&
  - \alpha \sum_{l=1}^D
  z_{s,l} \psi'(\alpha z_{s,l})
  \frac{\partial z_{s,l}}{\partial \eta_{s,k}}.
\end{align*}
The last equality is obtained from \eqref{eq:E_logy} and \eqref{eq:appB_sum}.
Symmetry gives that
$\mv{\mathcal{I}}_{\mv{\theta},\alpha} = \mv{\mathcal{I}}_{\alpha,\mv{\theta}}^\trsp$.

The last block of \eqref{eq:app_FI} is
\begin{align*}
  \mv{\mathcal{I}}_{\alpha,\alpha} =& -\pE\!\left(
    \frac{\partial^2 l(\mv{X}, \mv{\beta}, \alpha|\mv{y})}{\partial \alpha^2}
    \,\middle|\, \bullet \right)
  \\ =&
  -\left( \sum_{s=1}^{N_o} \psi'(\alpha)
    - \sum_{s=1}^{N_o} \sum_{k=1}^D z_{s,k}^2 \psi'(\alpha z_{s,k})
    - \frac{a_\alpha-1}{\alpha^2} \right).
\end{align*}

\section{The Posterior $\kappa|\mv{X}$}
\label{app:integral}
The posterior of $\kappa|\mv{X}$ is obtained by integrating out $\mv{\rho}$
from the joint posterior of $\kappa, \mv{\rho}|\mv{X}$.
With the densities for $\mv{X}$ and $\mv{\rho}$ given as
\begin{align}
  \mv{X}|\kappa, \mv{\rho} &\sim 
  \mathcal{N}\big(\mv{0},\mv{\rho}\otimes\mv{Q}^{-1}(\kappa)\big)
  & &\text{and} &
  \mv{\rho} &\sim IW(a_\rho\mathbb{I},b_\rho)
  \label{eq:app_X_rho}
\end{align}
in \eqref{eq:full} the posterior $\kappa|\mv{X}$ is
\begin{equation}
  \begin{split}
    \mathbb{P}(\kappa|\mv{X}) \propto& 
    \int  \mathbb{P}(\mv{X}|\kappa, \mv{\rho})
    \mathbb{P}(\kappa) \mathbb{P}(\mv{\rho}) d\mv{\rho}
    \\ \propto &
    \int 
    \det{\mv{\rho}^{-1}\otimes \mv{Q}(\kappa)}^{\frac{1}{2}}
    \exp \left( -\frac{1}{2} \mv{X}^\trsp \left(
        \mv{\rho}^{-1}\otimes \mv{Q}(\kappa) \right)
      \mv{X} \right)
    \mathbb{P}(\kappa) 
    \\ &
    \cdot \det{a_\rho\mathbb{I}}^{\frac{b_\rho}{2}}\det{\mv{\rho}}^{-\frac{b_\rho+d+1}{2}}
    \exp \left( -\frac{1}{2} \trace\left(\mv{\rho}^{-1} a_\rho\mathbb{I}\right) \right)
    \md\mv{\rho}.
  \end{split}
  \label{eq:app_integral1}
\end{equation}
Introducing vectorization such that $\vectorize(\mv{x}) = \mv{X}$,
where $\mv{x}=(X_1,\cdots, X_d)$ is a $N\times d$-matrix version of the
column-vector $\mv{X}$, the exponential term can be rewritten as
\begin{align*}
  & -\frac{1}{2} \mv{X}^\trsp \left(
    \mv{\rho}^{-1}\otimes \mv{Q}(\kappa) \right)
  \mv{X}
  =
  -\frac{1}{2} \mv{X}^\trsp \vectorize\left(
    \mv{Q}(\kappa)^\trsp \mv{x} \mv{\rho}^{-1} \right)
  \\ =&
  -\frac{1}{2} \trace\left( \mv{x}^\trsp 
    \mv{Q}(\kappa) \mv{x} \mv{\rho}^{-1} \right)
  =
  -\frac{1}{2} \trace\left( \mv{\rho}^{-1} \mv{x}^\trsp 
    \mv{Q}(\kappa) \mv{x} \right).
\end{align*}
The posterior in \eqref{eq:app_integral1} now simplifies to
\begin{equation*}
  \mathbb{P}(\kappa|\mv{X}) \propto
  \mathbb{P}(\kappa) \det{a_\rho\mathbb{I}}^{\frac{b_\rho}{2}} \det{\mv{Q}(\kappa)}^{\frac{d}{2}}
  \int \det{\mv{\rho}}^{-\frac{N+b_\rho+d+1}{2}}
  \exp \left( -\frac{1}{2} \trace\left(\mv{\rho}^{-1} 
      \left(a_\rho\mathbb{I} + \mv{x}^\trsp \mv{Q}(\kappa) \mv{x} \right)
    \right) \right) \md\mv{\rho}.
\end{equation*}
Recognizing the density of an unnormalized inverse-Wishart distribution under
the integral sign we normalise and obtain the posteriors
\begin{align*}
  \mv{\rho}|\kappa,\mv{X} &\sim
  IW\left(a_\rho\mathbb{I} + \mv{x}^\trsp \mv{Q}(\kappa) \mv{x}, b_\rho+N\right),
  \\
  \mathbb{P}(\kappa|\mv{X}) &\propto 
  \mathbb{P}(\kappa) \cdot 
  \frac{ a_\rho^{\frac{db_\rho}{2}} \det{\mv{Q}(\kappa)}^{\frac{d}{2}}}{ 
    \det{a_\rho\mathbb{I} + \mv{x}^\trsp \mv{Q}(\kappa) \mv{x}}^{\frac{N+b_\rho}{2}}}.
\end{align*}

\section{Parameter Estimates}
\label{app:para-est}

\newcommand{\captionTextParEst}{Parameter estimates (Est) and $95\%$ quantile
  (CI) for the two models (Full --- spatial model and RM --- regression model) used to reconstruct past land-cover composition from the PbLCC data.}

\begin{table}[H]
\caption{ \captionTextParEst }
\label{tab:para1700}
\centerline{
\begin{tabular}{crlrl}
\multicolumn{5}{l}{1700 CE}\\
\hline
\multicolumn{1}{c}{}&\multicolumn{2}{c}{Full}&\multicolumn{2}{c}{RM}\\
\cline{2-5}
\multicolumn{1}{c}{Parameter}&\multicolumn{1}{c}{$\text{Est}$}&\multicolumn{1}{c}{(CI)}&\multicolumn{1}{c}{$\text{Est}$}&\multicolumn{1}{c}{(CI)}\\
\hline
$\alpha$	&	9.55	&	(	7.64	,	12.93	)	&	6.07	&	(	5.31	,	6.86	)	\\
$\kappa$	&	0.23	&	(	0.12	,	0.39	)	&	-	&			-			\\
$\rho_{11}$	&	0.50	&	(	0.13	,	1.83	)	&	-	&			-			\\
$\rho_{12}$	&	0.23	&	(	0.01	,	1.11	)	&	-	&			-			\\
$\rho_{22}$	&	0.25	&	(	0.07	,	0.90	)	&	-	&			-			\\
$\beta_{10}$	&	-0.72	&	(	-1.85	,	0.23	)	&	-0.13	&	(	-0.24	,	0.00	)	\\
$\beta_{11}$	&	0.17	&	(	0.08	,	0.26	)	&	0.28	&	(	0.25	,	0.31	)	\\
$\beta_{12}$	&	-0.01	&	(	-0.12	,	0.10	)	&	-0.08	&	(	-0.12	,	-0.03	)	\\
$\beta_{13}$	&	-0.01	&	(	-0.20	,	0.18	)	&	-0.14	&	(	-0.24	,	-0.05	)	\\
$\beta_{20}$	&	-0.74	&	(	-1.56	,	0.02	)	&	-0.35	&	(	-0.49	,	-0.23	)	\\
$\beta_{21}$	&	0.07	&	(	-0.01	,	0.15	)	&	0.13	&	(	0.11	,	0.16	)	\\
$\beta_{22}$	&	-0.05	&	(	-0.13	,	0.03	)	&	-0.08	&	(	-0.12	,	-0.04	)	\\
$\beta_{23}$	&	-0.20	&	(	-0.38	,	-0.03	)	&	-0.30	&	(	-0.40	,	-0.20	)	\\
\hline
\end{tabular}}
\end{table}

\begin{table}[H]
\caption{ \captionTextParEst }
\label{tab:para1400}
\centerline{
\begin{tabular}{crlrl}
\multicolumn{5}{l}{1400 CE}\\
\hline
\multicolumn{1}{c}{}&\multicolumn{2}{c}{Full}&\multicolumn{2}{c}{RM}\\
\cline{2-5}
\multicolumn{1}{c}{Parameter}&\multicolumn{1}{c}{$\text{Est}$}&\multicolumn{1}{c}{(CI)}&\multicolumn{1}{c}{$\text{Est}$}&\multicolumn{1}{c}{(CI)}\\
\hline
$\alpha$	&	8.75	&	(	7.22	,	10.76	)	&	5.18	&	(	4.57	,	5.83	)	\\
$\kappa$	&	0.18	&	(	0.08	,	0.31	)	&	-	&			-			\\
$\rho_{11}$	&	0.37	&	(	0.10	,	0.98	)	&	-	&			-			\\
$\rho_{12}$	&	0.12	&	(	-0.02	,	0.47	)	&	-	&			-			\\
$\rho_{22}$	&	0.17	&	(	0.06	,	0.44	)	&	-	&			-			\\
$\beta_{10}$	&	-0.70	&	(	-2.34	,	0.77	)	&	-0.09	&	(	-0.21	,	0.03	)	\\
$\beta_{11}$	&	0.16	&	(	0.05	,	0.26	)	&	0.28	&	(	0.26	,	0.31	)	\\
$\beta_{12}$	&	0.02	&	(	-0.10	,	0.13	)	&	-0.07	&	(	-0.12	,	-0.02	)	\\
$\beta_{13}$	&	0.09	&	(	-0.10	,	0.28	)	&	-0.09	&	(	-0.19	,	0.00	)	\\
$\beta_{20}$	&	-0.56	&	(	-1.78	,	0.51	)	&	-0.15	&	(	-0.28	,	-0.03	)	\\
$\beta_{21}$	&	0.07	&	(	-0.03	,	0.16	)	&	0.14	&	(	0.11	,	0.17	)	\\
$\beta_{22}$	&	-0.03	&	(	-0.12	,	0.06	)	&	-0.07	&	(	-0.11	,	-0.03	)	\\
$\beta_{23}$	&	-0.18	&	(	-0.36	,	-0.01	)	&	-0.32	&	(	-0.42	,	-0.22	)	\\
\hline
\end{tabular}}
\end{table}

\begin{table}[H]
\caption{ \captionTextParEst }
\label{tab:para1000}
\centerline{
\begin{tabular}{crlrl}
\multicolumn{5}{l}{1000 BCE}\\
\hline
\multicolumn{1}{c}{}&\multicolumn{2}{c}{Full}&\multicolumn{2}{c}{RM}\\
\cline{2-5}
\multicolumn{1}{c}{Parameter}&\multicolumn{1}{c}{$\text{Est}$}&\multicolumn{1}{c}{(CI)}&\multicolumn{1}{c}{$\text{Est}$}&\multicolumn{1}{c}{(CI)}\\
\hline
$\alpha$	&	7.02	&	(	5.89	,	8.37	)	&	4.42	&	(	3.91	,	4.96	)	\\
$\kappa$	&	0.19	&	(	0.08	,	0.30	)	&	-	&			-			\\
$\rho_{11}$	&	0.32	&	(	0.10	,	0.80	)	&	-	&			-			\\
$\rho_{12}$	&	0.07	&	(	-0.04	,	0.27	)	&	-	&			-			\\
$\rho_{22}$	&	0.16	&	(	0.06	,	0.37	)	&	-	&			-			\\
$\beta_{10}$	&	0.19	&	(	-1.40	,	1.57	)	&	0.50	&	(	0.37	,	0.63	)	\\
$\beta_{11}$	&	0.24	&	(	0.13	,	0.35	)	&	0.30	&	(	0.27	,	0.33	)	\\
$\beta_{12}$	&	-0.03	&	(	-0.17	,	0.10	)	&	0.05	&	(	-0.01	,	0.11	)	\\
$\beta_{13}$	&	0.15	&	(	-0.06	,	0.37	)	&	-0.02	&	(	-0.12	,	0.09	)	\\
$\beta_{20}$	&	0.19	&	(	-1.07	,	1.21	)	&	0.55	&	(	0.43	,	0.68	)	\\
$\beta_{21}$	&	0.07	&	(	-0.01	,	0.16	)	&	0.12	&	(	0.09	,	0.14	)	\\
$\beta_{22}$	&	0.03	&	(	-0.08	,	0.13	)	&	0.02	&	(	-0.03	,	0.07	)	\\
$\beta_{23}$	&	-0.02	&	(	-0.20	,	0.18	)	&	-0.11	&	(	-0.22	,	-0.01	)	\\
\hline
\end{tabular}}
\end{table}

\begin{table}[H]
\caption{ \captionTextParEst }
\label{tab:para4000}
\centerline{
\begin{tabular}{crlrl}
\multicolumn{5}{l}{4000 BCE}\\
\hline
\multicolumn{1}{c}{}&\multicolumn{2}{c}{Full}&\multicolumn{2}{c}{RM}\\
\cline{2-5}
\multicolumn{1}{c}{Parameter}&\multicolumn{1}{c}{$\text{Est}$}&\multicolumn{1}{c}{(CI)}&\multicolumn{1}{c}{$\text{Est}$}&\multicolumn{1}{c}{(CI)}\\
\hline
$\alpha$	&	7.58	&	(	6.26	,	9.84	)	&	5.36	&	(	4.72	,	6.02	)	\\
$\kappa$	&	0.20	&	(	0.10	,	0.32	)	&	-	&			-			\\
$\rho_{11}$	&	0.21	&	(	0.07	,	0.69	)	&	-	&			-			\\
$\rho_{12}$	&	0.10	&	(	-0.02	,	0.64	)	&	-	&			-			\\
$\rho_{22}$	&	0.24	&	(	0.06	,	1.03	)	&	-	&			-			\\
$\beta_{10}$	&	0.41	&	(	-0.70	,	1.48	)	&	0.38	&	(	0.24	,	0.53	)	\\
$\beta_{11}$	&	0.23	&	(	0.13	,	0.33	)	&	0.23	&	(	0.20	,	0.26	)	\\
$\beta_{12}$	&	-0.19	&	(	-0.36	,	-0.03	)	&	-0.04	&	(	-0.11	,	0.03	)	\\
$\beta_{13}$	&	0.04	&	(	-0.17	,	0.25	)	&	-0.04	&	(	-0.14	,	0.07	)	\\
$\beta_{20}$	&	0.61	&	(	-0.66	,	1.58	)	&	0.99	&	(	0.85	,	1.12	)	\\
$\beta_{21}$	&	0.01	&	(	-0.09	,	0.10	)	&	0.08	&	(	0.06	,	0.11	)	\\
$\beta_{22}$	&	-0.01	&	(	-0.16	,	0.14	)	&	0.00	&	(	-0.06	,	0.05	)	\\
$\beta_{23}$	&	-0.05	&	(	-0.24	,	0.15	)	&	-0.25	&	(	-0.35	,	-0.15	)	\\
\hline
\end{tabular}}
\end{table}

\section{Maps of Estimated Land Cover}
\label{app:results}

\newcommand{\captionTextMaps}[1]{Results for the #1 time period: the first
  row shows the PbLCC data, and the other rows show the reconstructions for the
  full spatial model (Full) and the regression model (RM).}

\begin{figure}[H]
\noindent\makebox[\textwidth]{
\includegraphics[scale=0.6]{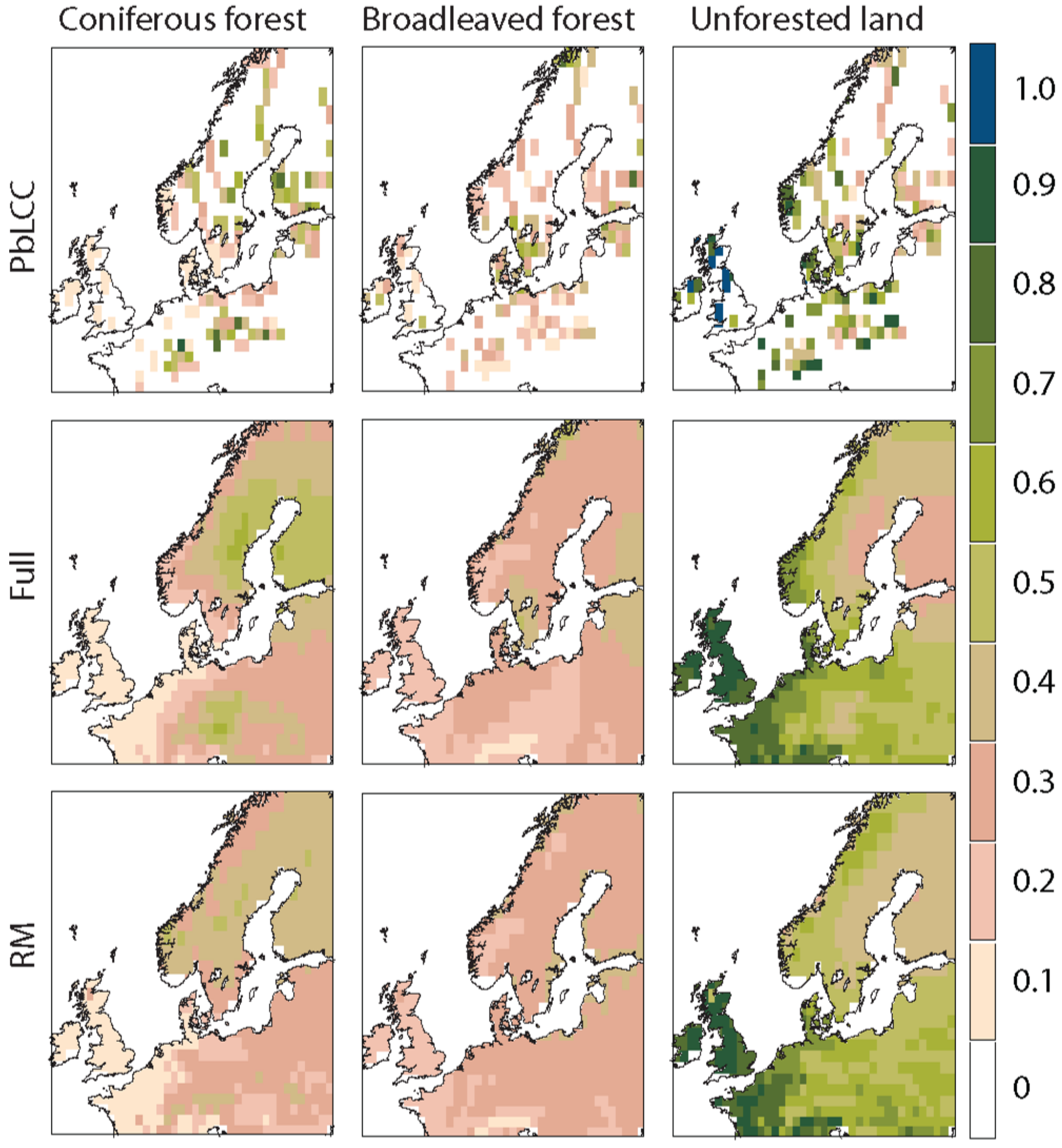}}
\caption{ \captionTextMaps{1725 CE} }
\label{fig:1725}
\end{figure}
\begin{figure}[H]
\noindent\makebox[\textwidth]{
\includegraphics[scale=0.6]{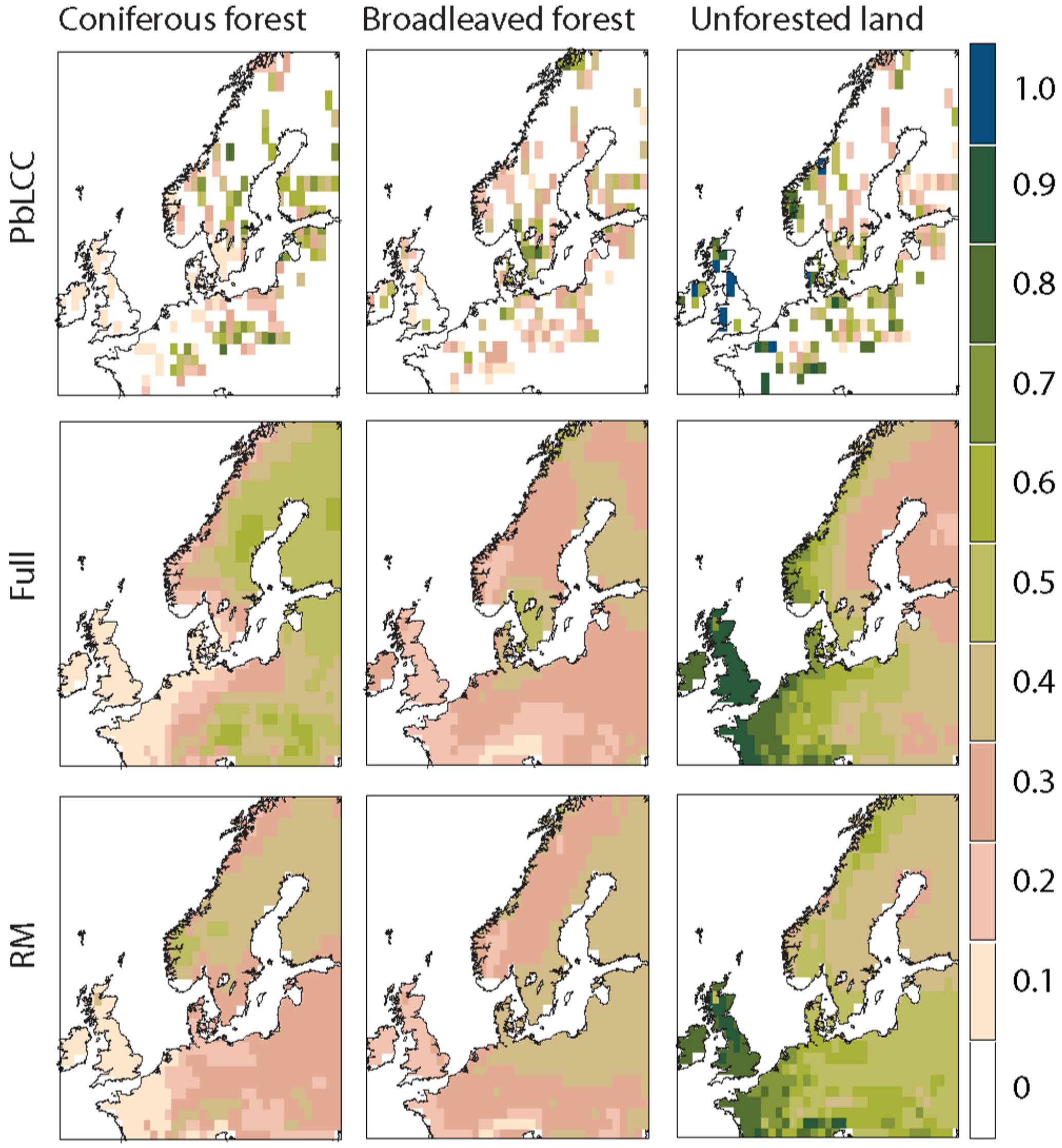}}
\caption{ \captionTextMaps{1425 CE} }
\label{fig:1425}
\end{figure}
\begin{figure}[H]
\noindent\makebox[\textwidth]{
\includegraphics[scale=0.6]{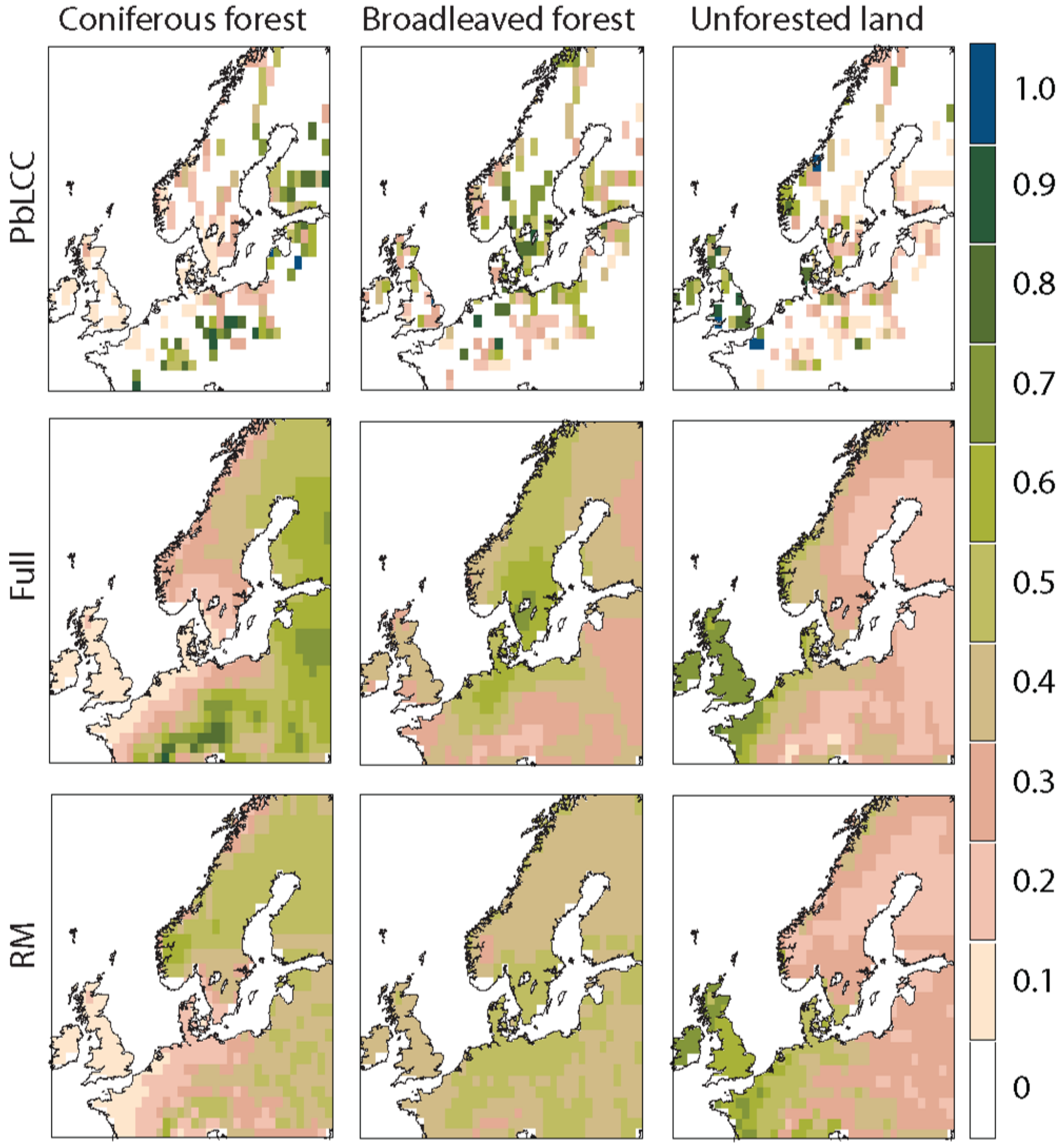}}
\caption{ \captionTextMaps{1000 BCE} }
\label{fig:1000}
\end{figure}
\begin{figure}[H]
\noindent\makebox[\textwidth]{
\includegraphics[scale=0.6]{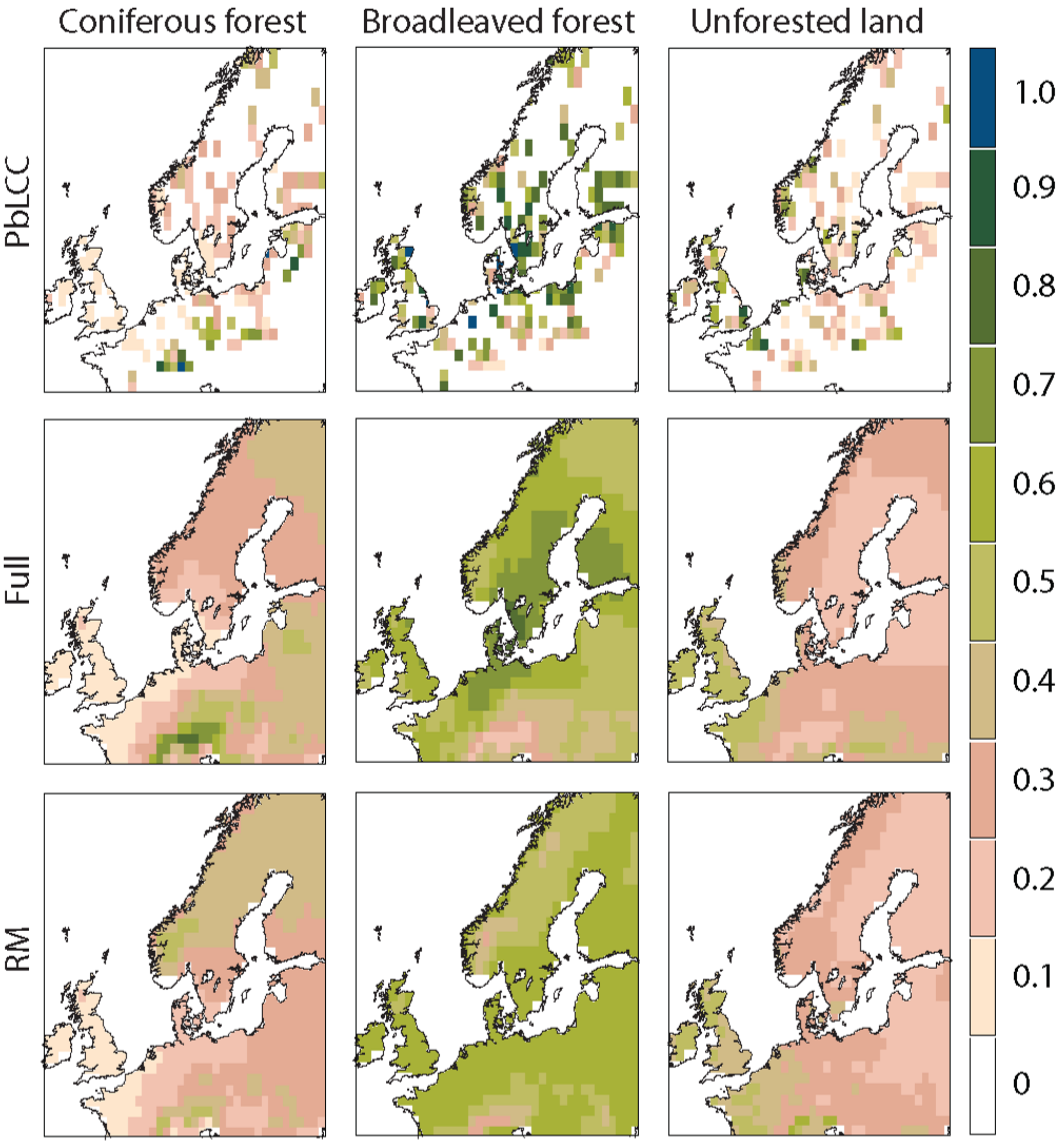}}
\caption{ \captionTextMaps{4000 BCE} }
\label{fig:4000}
\end{figure}

\section{Uncertainties in Estimated Land Cover}
\label{app:results_CI}

\newcommand{\captionTextMapsCI}[1]{The first column shows the reconstructed land-cover composition for the #1 time period, using the full spatial model. Columns 2 and 3, row 1 (with thick/red axes), show the maximum and minimum of $95\%$ elliptical confidence regions for Coniferous; rows 2 and 3 give the corresponding Broadleaved and Unforested compositions. Columns 4 and 5 (row 2 with thick/red axes) gives the bounds for the Broadleaved composition while columns 6 and 7 show the bounds for Unforested land (row 3 with thick/red axes). The concept of joint confidence interval for compositions is shown in Fig \ref{fig:ellipse}.}

\subsection{Maps of Uncertainties}
\begin{figure}[H]
\noindent\makebox[\textwidth]{
\includegraphics[scale=0.6]{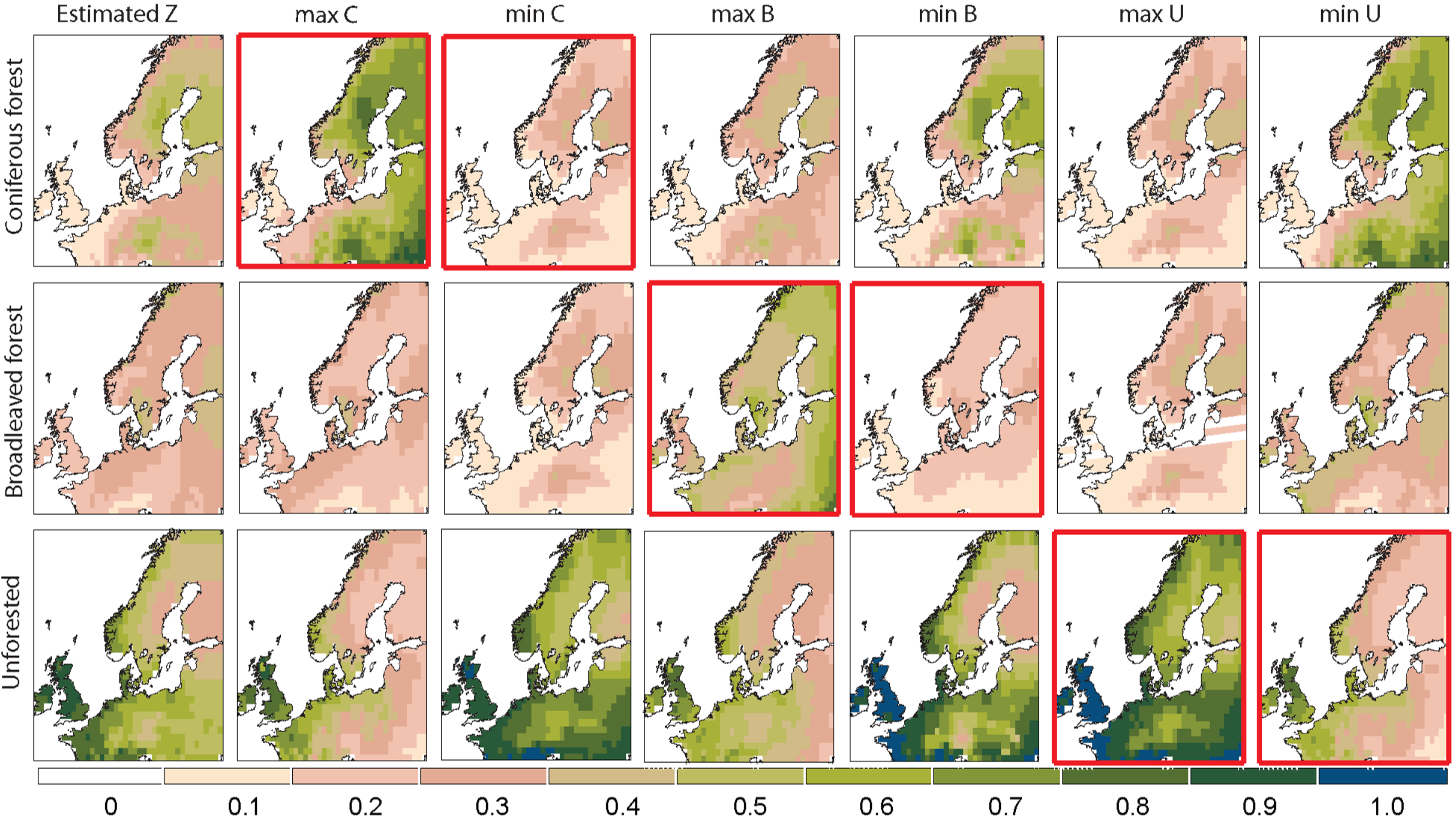}}
\caption{ \captionTextMapsCI{1725 CE} }
\label{fig:results_CI_AD1725}
\end{figure}
\begin{figure}[H]
\noindent\makebox[\textwidth]{
\includegraphics[scale=0.65]{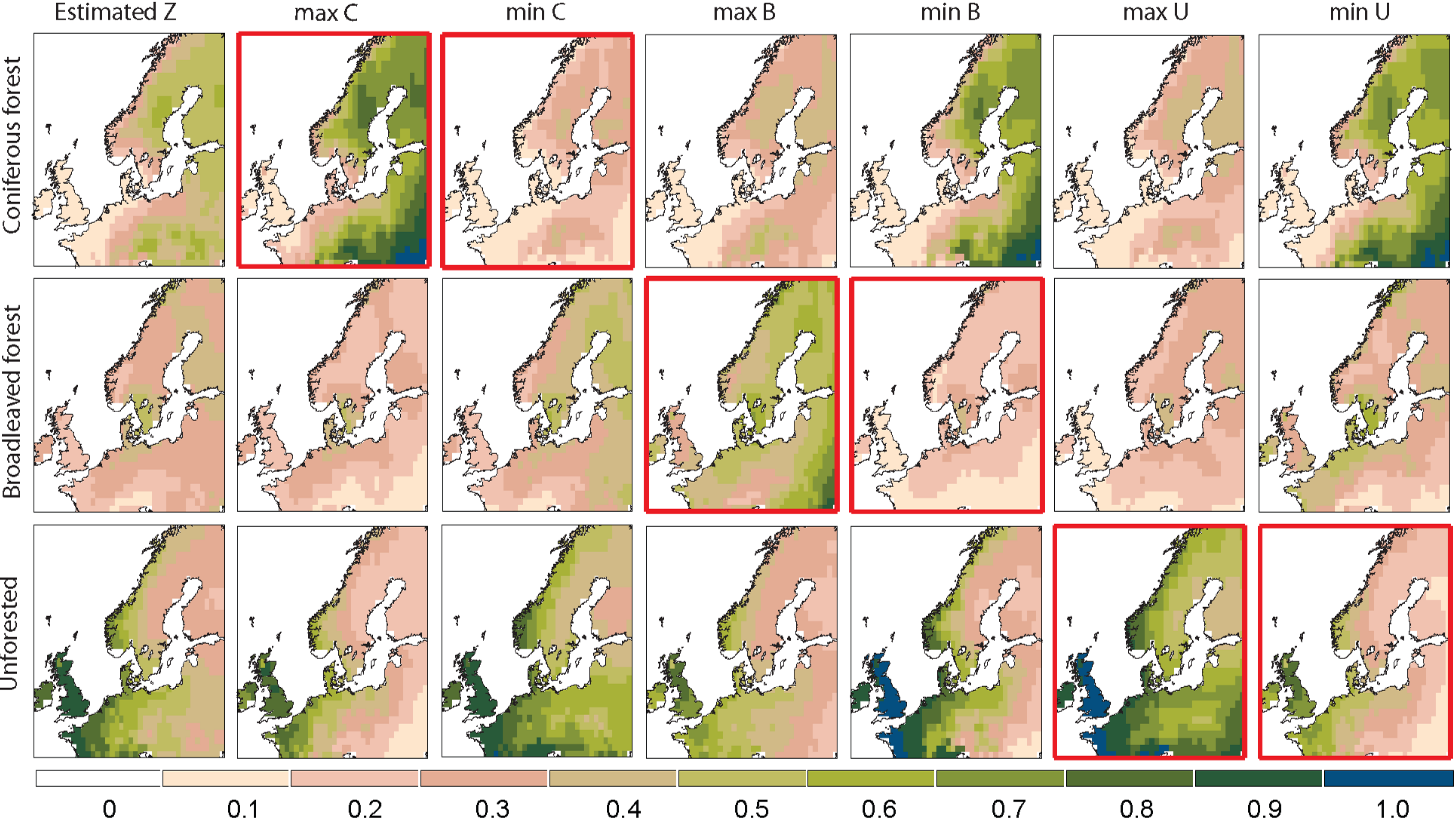}}
\caption{ \captionTextMapsCI{1425 CE} }
\label{fig:results_CI_AD1425}
\end{figure}
\begin{figure}[H]
\noindent\makebox[\textwidth]{
\includegraphics[scale=0.65]{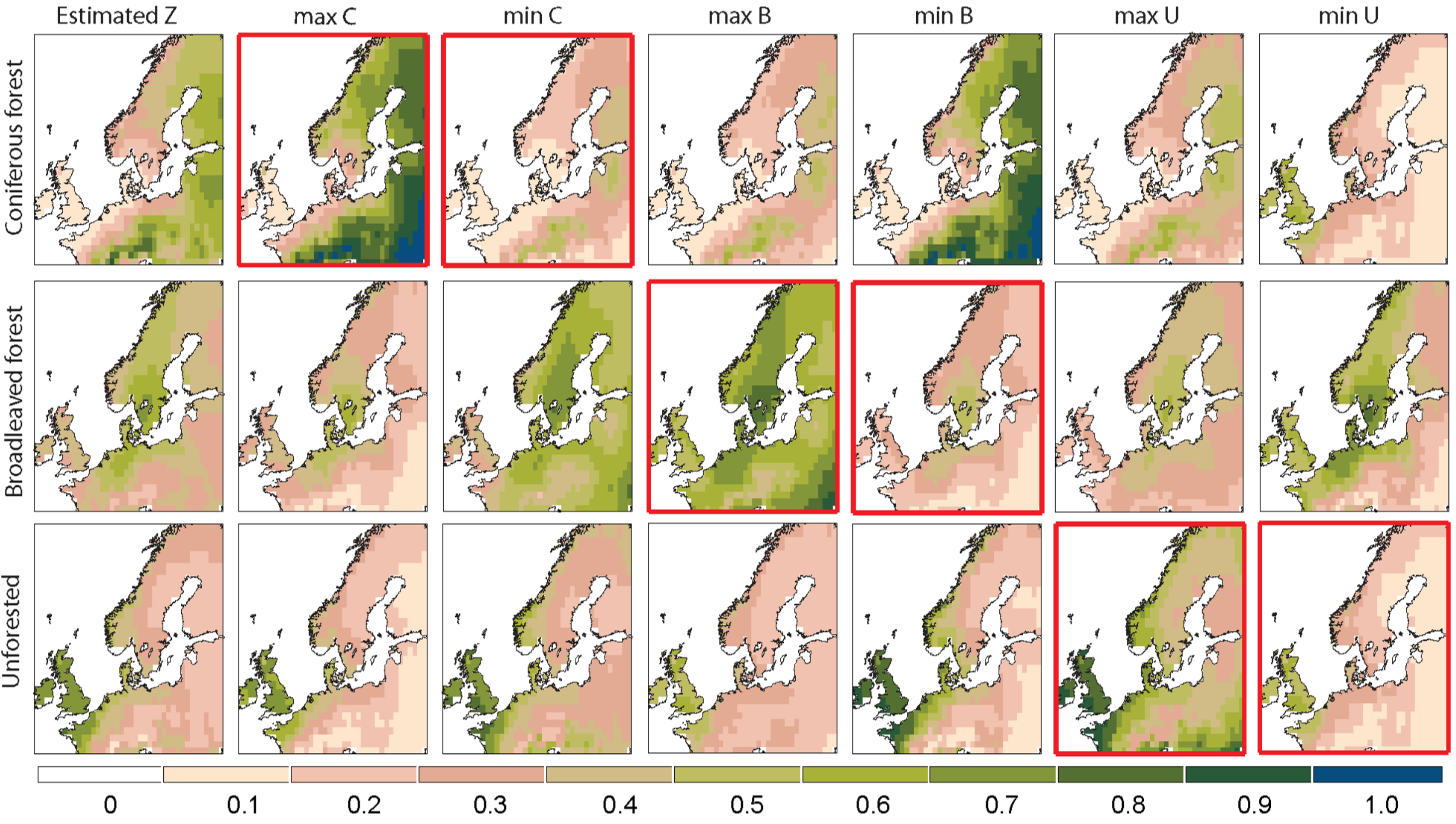}}
\caption{ \captionTextMapsCI{1000 BCE} }
\label{fig:results_CI_BC1000}
\end{figure}
\begin{figure}[H]
\noindent\makebox[\textwidth]{
\includegraphics[scale=0.65]{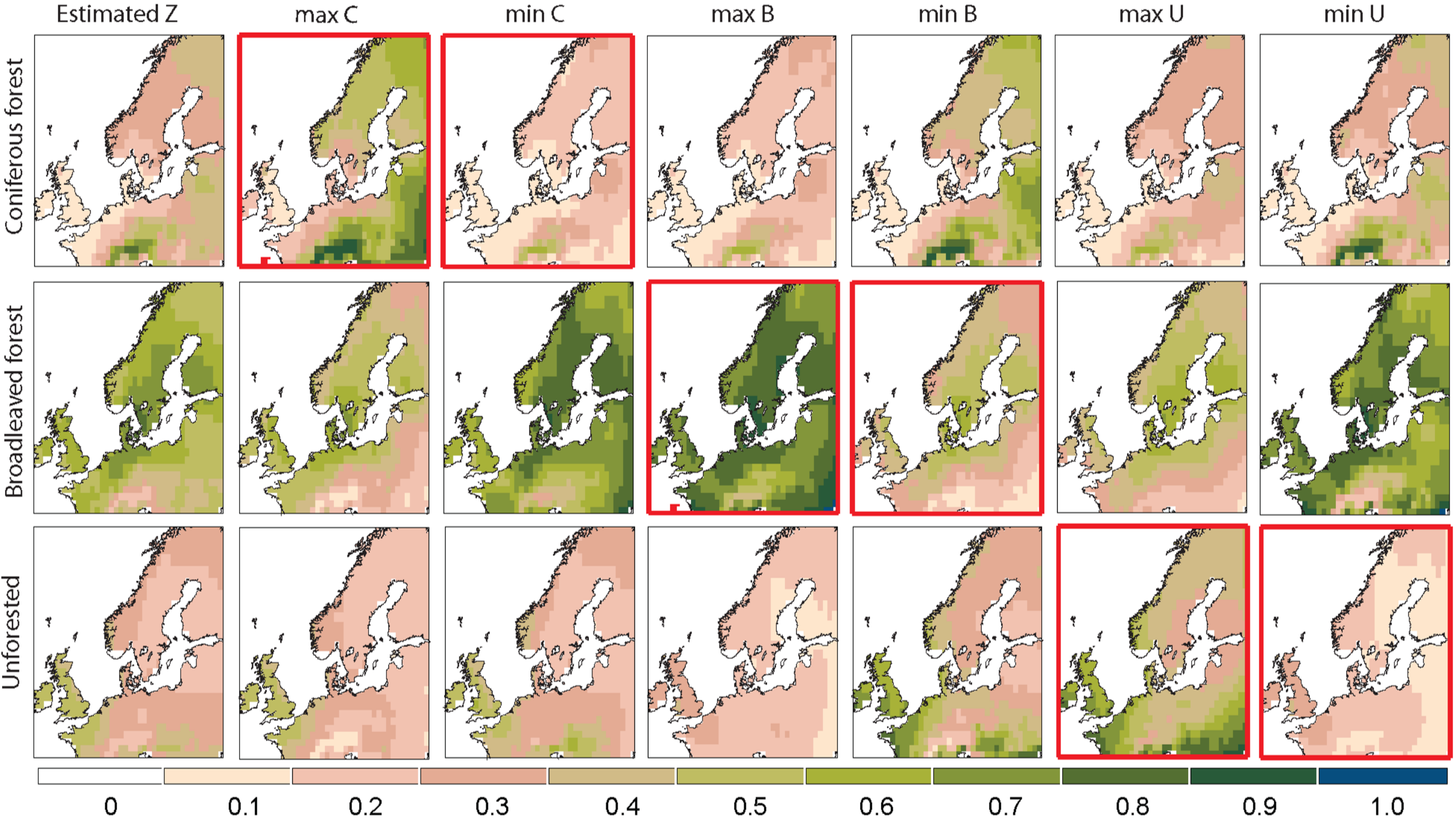}}
\caption{ \captionTextMapsCI{4000 BCE} }
\label{fig:results_CI_BC4000}
\end{figure}

\subsection{Confidence Regions for Selected Locations}
\newcommand{\captionTextPRCR}[1]{The first row shows the locations of the three selected grid cells for the #1 time period. The second row shows the ternary confidence regions along with the reconstructions for the two models (Full---spatial model; RM---regression model) and the values of the PbLCC data and the LPJ-GUESS$_{\text{KK}}$ land cover covariate at each location. The third row shows the ternary prediction regions and the same values.}

\begin{figure}[H]
\noindent\makebox[\textwidth]{
\includegraphics[scale=0.6]{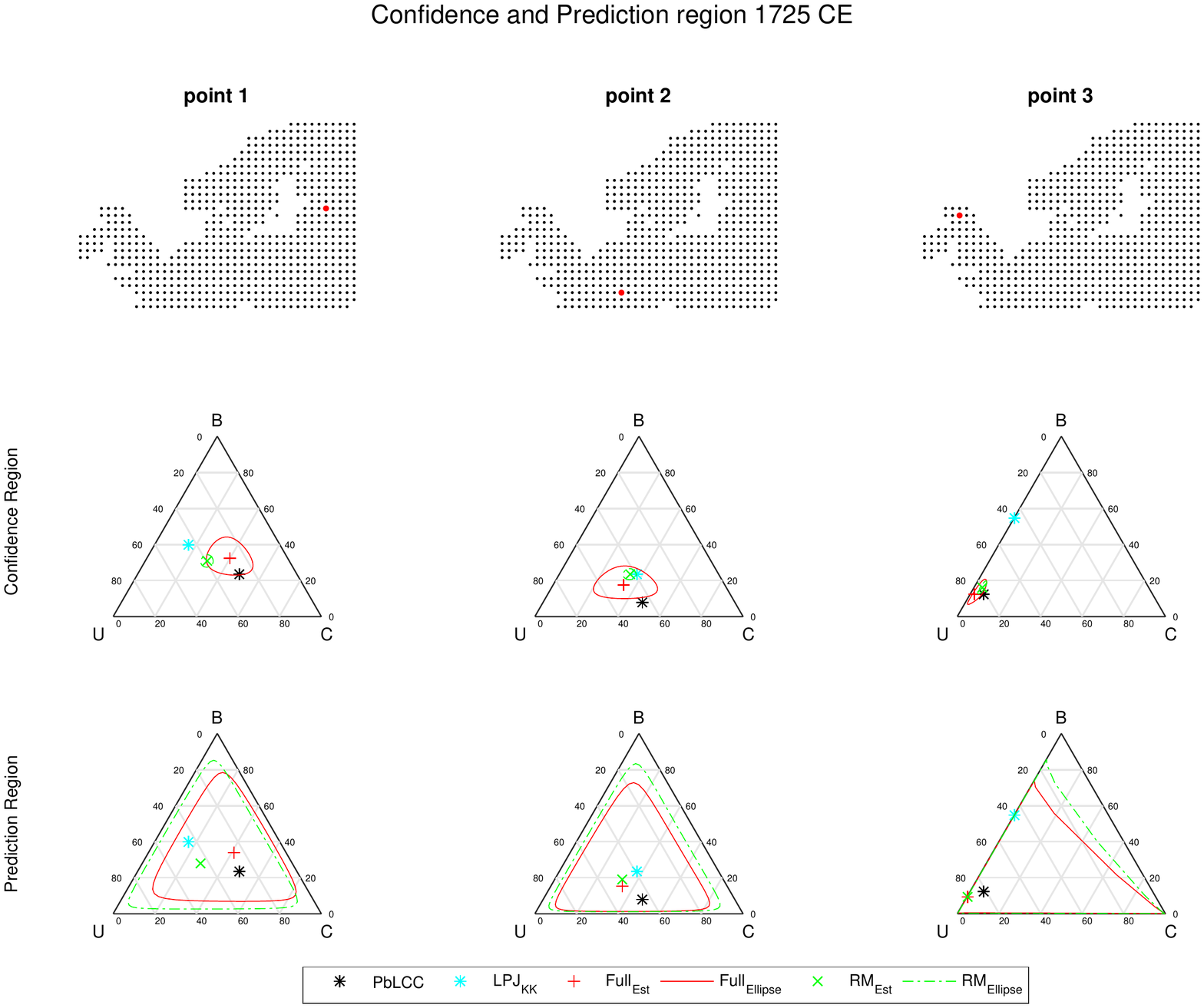}}
\caption{ \captionTextPRCR{1725 CE} }
\label{fig:PR_CR_AD1725}
\end{figure}

\begin{figure}[H]
\noindent\makebox[\textwidth]{
\includegraphics[scale=0.7]{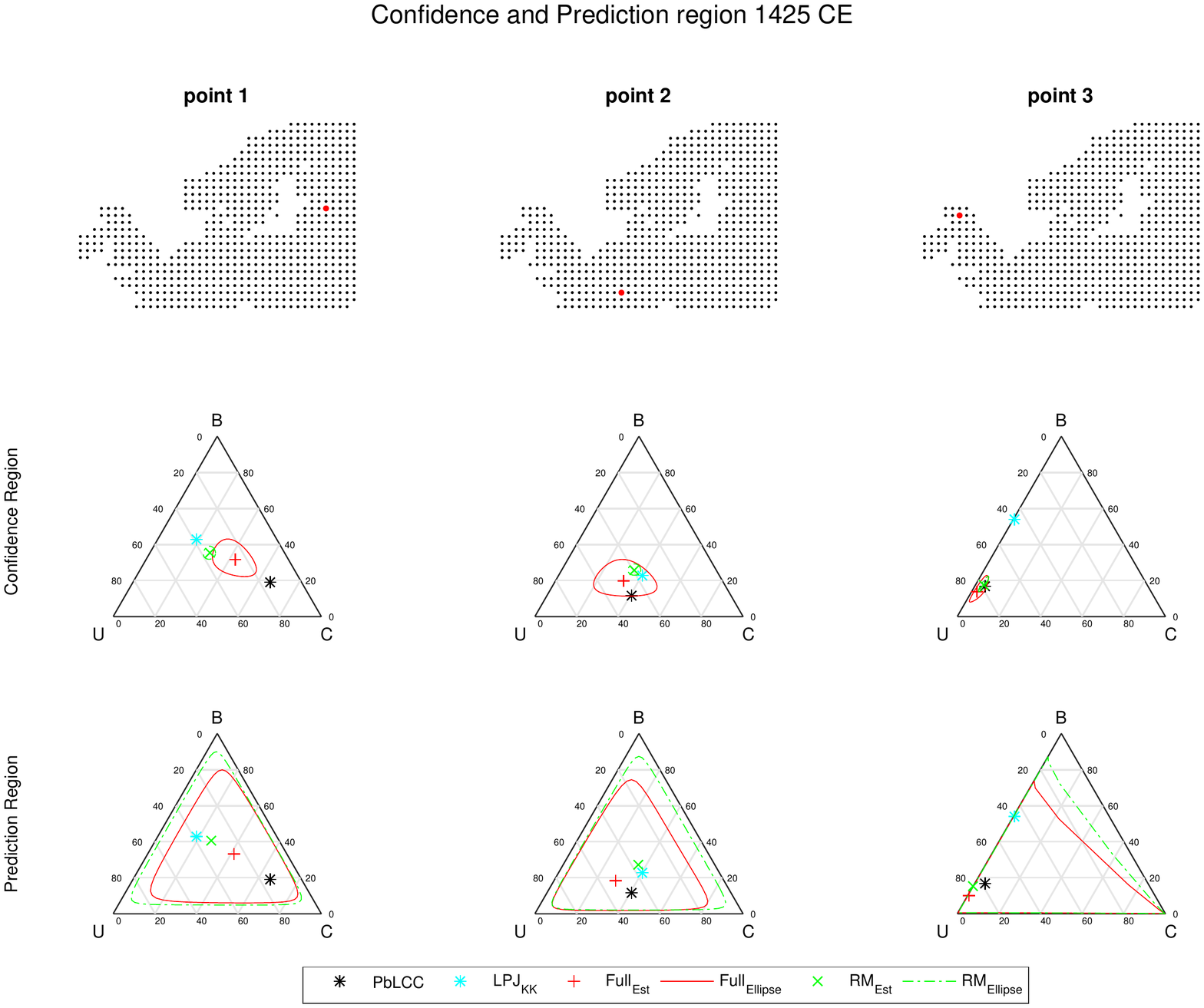}}
\caption{ \captionTextPRCR{1425 CE} }
\label{fig:PR_CR_AD1425}
\end{figure}

\begin{figure}[H]
\noindent\makebox[\textwidth]{
\includegraphics[scale=0.7]{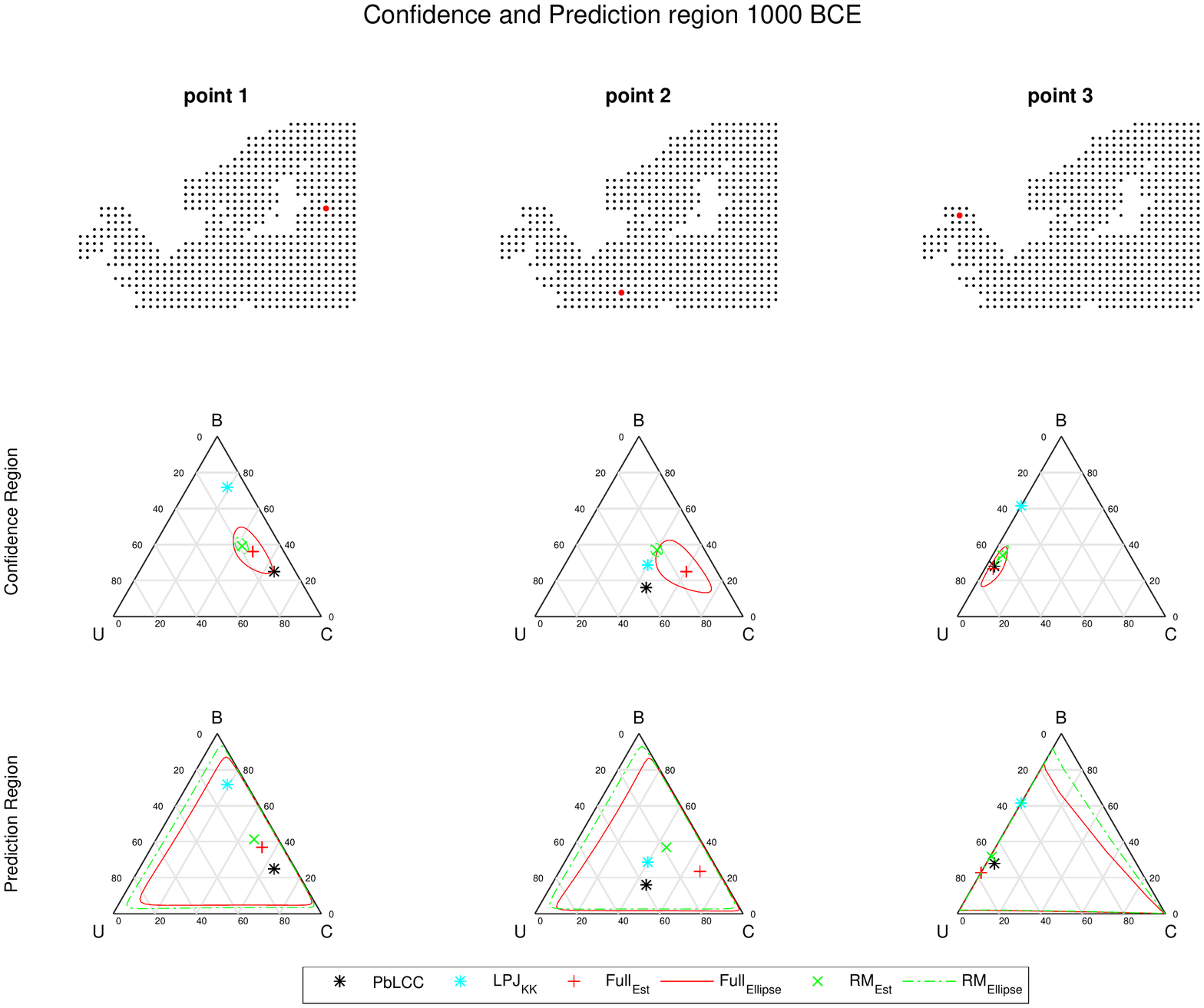}}
\caption{ \captionTextPRCR{1000 BCE} }
\label{fig:PR_CR_BC1000}
\end{figure}

\begin{figure}[H]
\noindent\makebox[\textwidth]{
\includegraphics[scale=0.7]{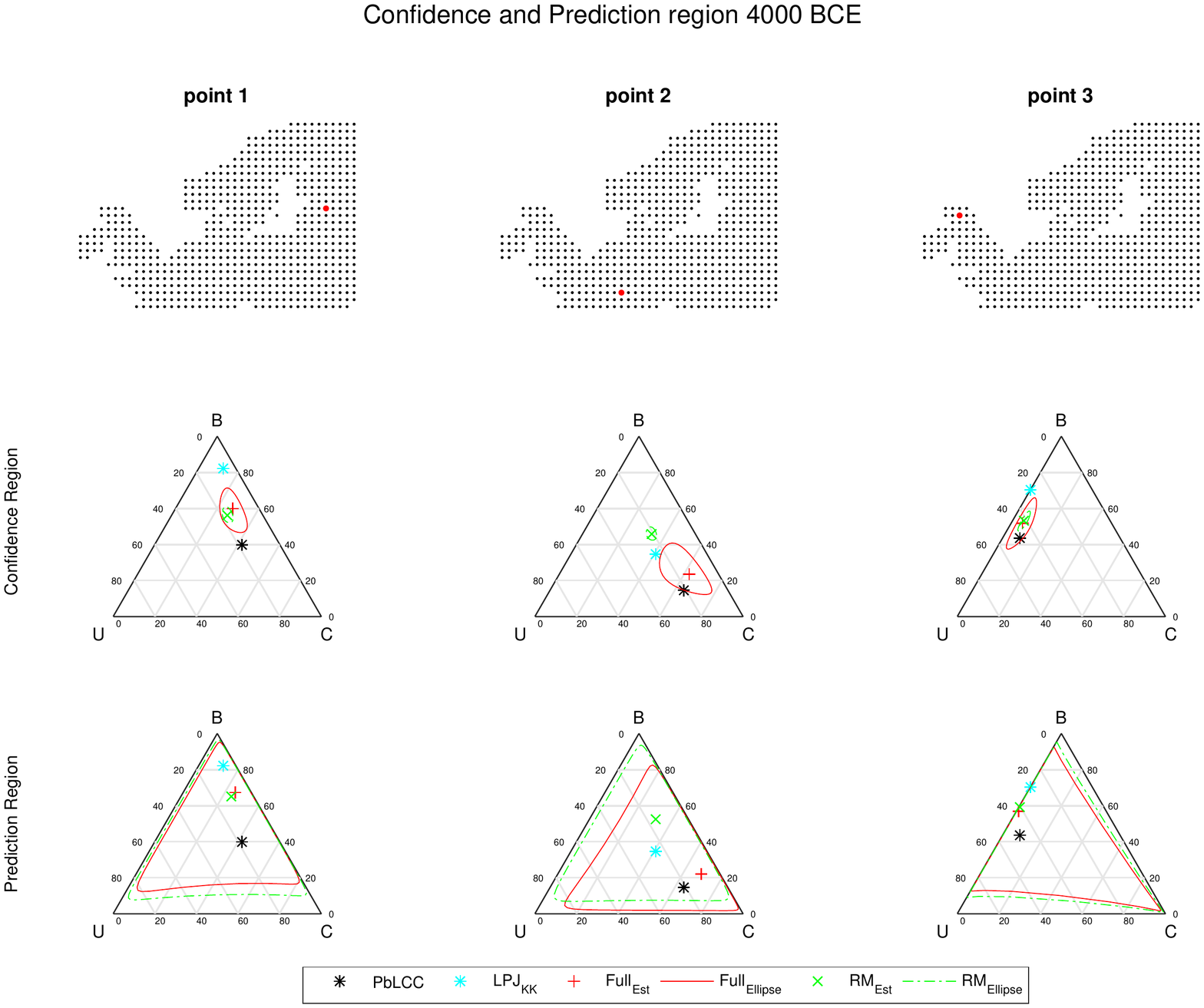}}
\caption{ \captionTextPRCR{4000 BCE} }
\label{fig:PR_CR_BC4000}
\end{figure}

\end{document}